\documentclass{article}

\usepackage{PRIMEarxiv}

\usepackage[utf8]{inputenc} 
\usepackage[T1]{fontenc}    
\usepackage{hyperref}       
\usepackage{url}            
\usepackage{booktabs} 
\usepackage{bm}
\usepackage[normalem]{ulem}

\usepackage{amsfonts}       
\usepackage{nicefrac}       
\usepackage{microtype}      
\usepackage{fancyhdr}       
\usepackage{graphicx}       
\usepackage{amssymb,amsfonts,amsmath,latexsym,dsfont}
\usepackage{makecell}       
\usepackage{array}
\usepackage{tikz,graphics,color,fullpage,float,epsf,caption,subcaption}

\title{Complex Valued Deep Operator Network (DeepONet) $[\mathcal{G}]$ for Three Dimensional Maxwell's Equations: $\mathcal{G} \in \mathbb{C}^{m \times n}$}

\author{
      Qile Jiang \\
      Division of Applied Mathematics\\
      Brown University \\
      Providence, RI 02906\\
      \texttt{qile\_jiang@brown.edu} \\
\And
      Marc Salvadori \\
      Hypercomp Inc. \\
      Westlake Village, CA 91362 \\
      \texttt{msalvadori13@hypercomp.net} \\
\And
      Dale Ota \\
      Hypercomp Inc. \\
      Westlake Village, CA 91362 \\
      \texttt{dkota@hypercomp.net} \\
\And
      Vijaya Shankar \\
      Hypercomp Inc. \\
      Westlake Village, CA 91362 \\
      \texttt{vshankar@hypercomp.net} \\
\And
      Khemraj Shukla$^*$ \\
      Division of Applied Mathematics\\
      Brown University \\
       Providence, RI 02906\\
      \texttt{khemraj\_shukla@brown.edu} \\
}

\begin{document}
\maketitle

\begin{abstract}
\textcolor{black}{Maxwell's equations}, a system of linear partial differential equations (PDEs), describe the behavior of electric and magnetic fields in time and space and \textcolor{black}{are} essential for many important electromagnetic applications. Although numerical methods have been applied successfully in the past, the primary challenge in solving these equations arises from the frequency of electromagnetic fields, which depends on the shape and size of the objects to be resolved. The frequency dictates the grid size used for the spatial discretization of the PDEs. Since the domain of influence for these equations is compactly supported, even a small perturbation in frequency necessitates a new discretization of Maxwell's equations, resulting in substantial computational costs. In this work, we investigate the potential of neural operators, particularly the Deep Operator Network (DeepONet) and its variants, as a surrogate model for Maxwell's equations. Existing DeepONet implementations are restricted to real-valued data in $\mathbb{R}^n$, but since the time-harmonic Maxwell's equations yield solutions in the complex domain $\mathbb{C}^n$, a specialized architecture is required to handle complex algebra. We propose a formulation of DeepONet for complex data, define the forward pass in the complex domain, and adopt a reparametrized version of DeepONet for more efficient training. We also propose a unified framework to combine a plurality of DeepONets, trained for multiple electromagnetic field components, to incorporate the boundary condition. We conduct computational experiments on a three-dimensional metallic sphere without singularities and on a metallic almond-shaped target to demonstrate the effectiveness of the proposed method for problems involving singularity-prone solutions. As shown by computational experiments, our method significantly enhances the efficiency of predicting scattered fields from a spherical object at arbitrary high frequencies.
\end{abstract}



\section{Introduction:}\label{sec:intro}
Electromagnetics is a fundamental subject that studies electric and magnetic fields, their interactions, and their effects on materials and systems, which is crucial to a variety of daily and industrial applications, including wireless communication \cite{bondeson2012computational, nassa2011wireless, sarkar2006discussion}, radar \cite{zohuri2020radar, ernst2007full, bergmann1996numerical}, and medical imaging \cite{soni2005finite, afsari2018rapid}. Central to this field are Maxwell's equations, which describe how electromagnetic waves propagate and respond to external charges and currents. However, analytical solutions in closed form are known for only a very limited number of special and trivial cases. This limitation has motivated the development of various high order numerical methods \textcolor{black}{to solve} the Maxwell's equation in time and frequency domain \cite{hesthaven2002nodal, bondeson2012computational, maggio2004least}. However, because the grid resolution depends on the size of the object being resolved, the numerical methods can become computationally demanding due to the high frequencies and small skin depth required to resolve the small objects. To address this issue, we propose using a neural network-based approximate operator as a surrogate for the Maxwell's equation solver. 

Recently, computational problems in science and engineering have seen significant advancements through the application of deep learning methods. Commonly used neural network solvers can be broadly divided into two types: the first type focuses on learning mappings between input $\mathbb{R}^m$ and output data ${\mathbb{R}^m}$, while the second type, known as neural operators, \textcolor{black}{learns mappings} between two functional spaces. In this work, we focus on the latter—neural operators \cite{kovachki2024operator, Lu_DeepONet, boulle2023mathematical}—which aim to approximate unknown operators that are often linked to the  solution operator to differential equations. In mathematical terms, given data pairs $(v, u)$, where $v \in \mathcal{V}$ and $u \in \mathcal{U}$ belong to two function spaces, and a potentially nonlinear operator $\mathcal{G}: \mathcal{V} \mapsto \mathcal{U}$ such that $\mathcal{G}(v) = u$, the objective is to find a neural network approximation of $\mathcal{G}$, denoted by $\hat{\mathcal{G}}$, such that it can generalize to unseen data. By leveraging neural operators, we can efficiently replace highly complex and computationally intensive multiphysics systems and deliver functional outputs in real-time.

\begin{figure}[h!]
    \centering
    \includegraphics[width=0.8\textwidth]{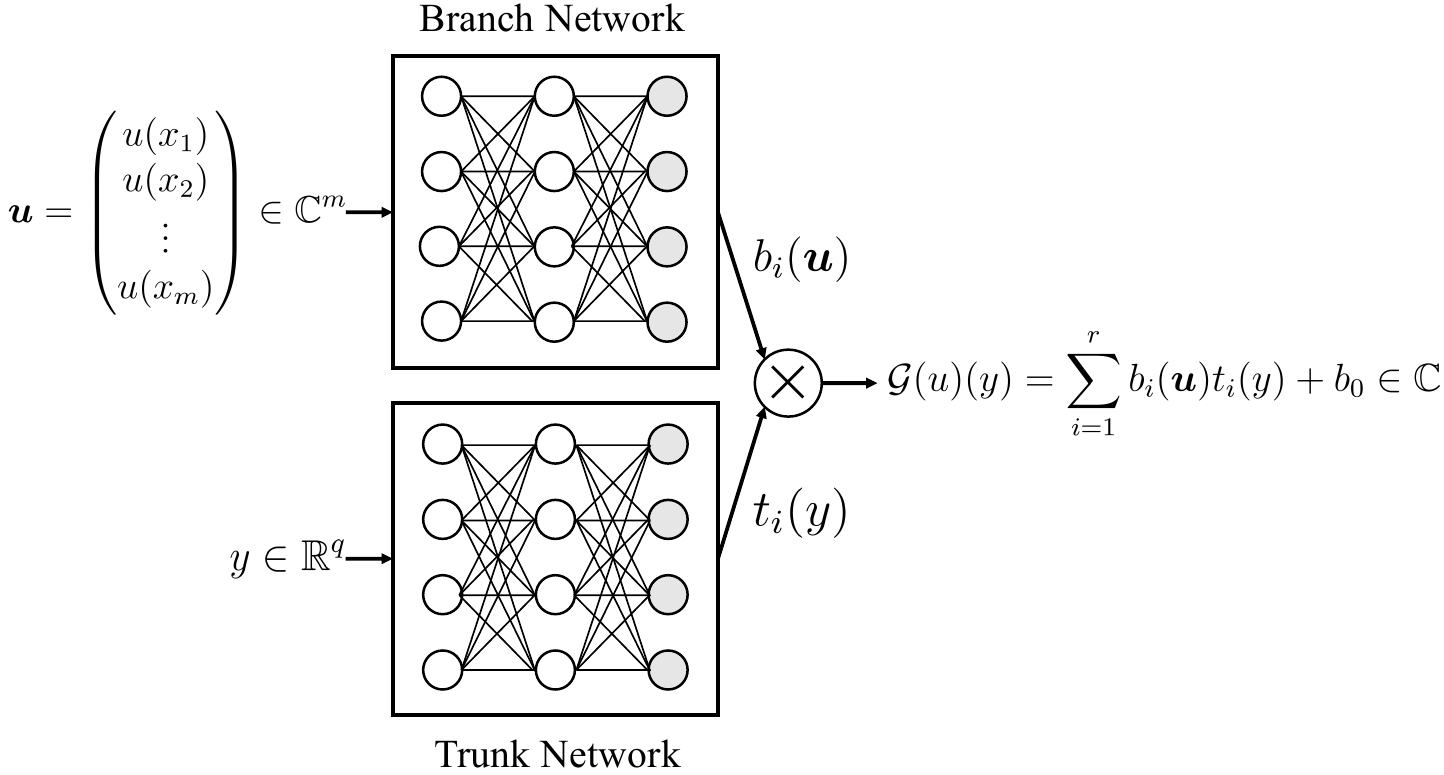}
    \caption{A schematic representation of a DeepONet, which is a neural operator trained to learn the mapping from the input function $u(x)$ to the output function $\mathcal{G}(u)(y)$, evaluated at $y \in \mathbb{R}^q$. DeepONet consists of two networks: a branch network, which processes the input function evaluations $ [ u(x_1), u(x_2), \dots, u(x_m)]^\top$, and a trunk network, which takes in the coordinates $y$.}
    \label{fig:DeepOnet diagram}
\end{figure}

Deep Operator Network (DeepONet) \cite{Lu_DeepONet} is a neural network based parametrized model for learning nonlinear operators and capturing the inherent relationships between input and output functions. DeepONets are built upon the universal operator approximation theorem \cite{chen_universal_thm_neural_operator, Lu_DeepONet}, which is an infinite dimensional analogue of the universal approximation theorem for neural networks \cite{cybenko1989approximation, hornik1991approximation}. \textcolor{black}{An} architectural diagram of a DeepONet with the commonly used nomenclature for its components is shown in \autoref{fig:DeepOnet diagram}. The DeepONet is a two-pronged deep learning network consisting of a branch network, which can take a multi-fidelity or multi-modal input \cite{de2023bi, lu2022multifidelity, howard2022multifidelity, zhu2023reliable}, and a trunk network \textcolor{black}{that encodes} the independent variables defining the output space, e.g., in space–time coordinates. To recover the solution of a PDE the output of branch network \textcolor{black}{is projected on} the output of trunk networks via an inner-product. DeepONet has seen success in solving complex problems in diverse scientific and engineering problems \cite{lin2021bubble, gao2024multiscale, SHUKLA2024107615, cai2021deepm, hao2023instability, shukla2024deep}, climate \cite{kissas2022learning, he2023hybrid, wang2024learning}, and material science \cite{goswami2022physics, he2023novel, lu2022multifidelity}. In \autoref{fig:DeepOnet diagram}, we have shown fully connected architecture but any architecture can be used depending on the dimensionality of the training and testing data.

All the existing applications of DeepONets focus on real-valued input and output data. However, many scientific problems involve complex-valued data, particularly in the solutions of important partial differential equations such as the Schrodinger equation \cite{berezin2012schrodinger, zwiebach2022schrodinger}, Helmholtz equation \cite{ernst2011difficult}, and Maxwell's equations \cite{munshi2022complexmaxwell, reggia2023complexmaxwell}. \textcolor{black}{Although there are studies of solving Maxwells equation in 2D using PINNs \cite{chen2020physics, wiecha2021deep, zhu2021fast, lim2022maxwellnet}, they consider the system with real valued solutions.} To address the challenge of representing both the magnitude ($m$) and phase ($\varphi$) of a complex number $z = m e^{i \varphi} \in \mathbb{C}$ and take advantage of the complex algebra, this work introduces the complex-valued DeepONet for learning operator mappings between complex-valued functions as input and output. Specifically, we demonstrate the effectiveness of using the complex-valued surrogate model for electromagnetic computations governed by the system of time-harmonic 3D Maxwell's equations, where the DeepONet is leveraged to learn the general mapping from incident fields to the scattered fields parametrized by an arbitrary angular frequency. 

To incorporate \textcolor{black}{the} complex valued input and output in a neural network requires redefining all the three steps (1) Affine Transform (2) Non-linearity and (3) backpropagation. The first step requires matrix-vector product, which is well defined for complex valued matrices. The second step will require defining the non-linear activation function $f:\mathbb{C}^n \to \mathbb{C}^m$ for complex valued variables as domain and co-domain. This will require defining the new complex valued activation function. However, there are existing definitions of such functions and in this study we use the definition proposed by \cite{kim2001complex}. \textcolor{black}{In the third step} we require computing the gradient of loss function with respect to complex parameters of neural network. To achieve this efficiently, we have used Wirtinger calculus \cite{amin2011wirtinger}. The usage of complex valued neural network is not common, but a few studies have shown great promise of using complex valued model for classification \cite{lowe2022complex} and regression \cite{virtue2017better, amin2011complex, nitta1997extension} problems.

This paper is structured as follows: In \autoref{sec:prob_setup}, we present the definition of the problem, governing equations, the geometry of the domain, and the data generation techniques used \textcolor{black}{to train the model}. Next, in \autoref{sec:Model}, we describe the architecture and forward passes for the complex-valued DeepONet. In addition, we \textcolor{black}{adapt} a two-step training approach to reparametrize the complex-valued network and demonstrate how separate DeepONets can be combined to enforce the boundary conditions for a perfect electric conductor (PEC). Lastly, in \autoref{Results}, we highlight the superior results of the complex-valued model compared to its real-valued counterpart, along with the post-processed radar cross-section (RCS) results.
{
\subsection{Contributions:}
\begin{enumerate}
\item In this work, we propose a composite neural network framework in which learning is carried out directly in the complex space $\mathbb{C}^N$. Unlike traditional approaches that split complex-valued solutions into their real and imaginary components and process them using separate real-valued neural networks, our method avoids such decomposition, which can obscure the inherent coupling between the two components and introduce unnecessary architectural complexity. Instead, we operate natively in the complex domain by employing complex-valued activation functions and developing a corresponding backpropagation algorithm defined entirely in complex space. This unified formulation preserves the intrinsic structure of complex-valued representations and enables more faithful and efficient learning for problems where the governing physics is naturally expressed in complex form.

\item We train six separate DeepONets to model the electric field ($\bm{E}$) and magnetic field ($\bm{H}$) components individually, with the goal of achieving improved representation and approximation accuracy for each component. However, in order to properly impose the underlying physics on the DeepONet predictions, it is necessary to consider all six components simultaneously. To address this challenge, we propose a novel approach that combines these six individual DeepONets into a unified framework, allowing us to enforce the physics of electromagnetic waves through the governing partial differential equations (PDEs). This strategy ensures that the DeepONet predictions remain physically consistent while benefiting from the specialized learning of each field component.

\item In this study, we also demonstrate the applicability of the proposed DeepONet to a three-dimensional almond-shaped complex geometry that gives rise to singularities. Furthermore, we address the training of DeepONets for complex geometries, such as the almond shape with two singularity locations, highlighting the model's capability to handle multiple singularities simultaneously.
\end{enumerate}
}

\section{Problem setup}\label{sec:prob_setup}
The computational domain considered here is a sphere with the surface being perfectly electrically conductive materials (PEC). In this section, we define the problem setup, governing equations, and numerical methods used to generate the data for the given 3D domain. 

\subsection{System of time-harmonic Maxwell's equations} \label{sec:equation_setup}

In this work, we devise the DeepONet to address the direct scattering problem, which is commonly encountered in various important electromagnetic applications and system is governed by Maxwell's equations, which reads as follows,
\begin{align}\label{eq:Maxwell3DEq}
-i \omega \varepsilon \boldsymbol{E} - \nabla \times \boldsymbol{H} &= 0 \quad \text{in } \Omega, \\
-i \omega \mu \boldsymbol{H} + \nabla \times \boldsymbol{E} &= 0 \quad \text{in } \Omega,
\end{align}
where $\boldsymbol{E}$ and $\boldsymbol{H}$ are the electric and magnetic fields, $\omega$ is the angular frequency of the radiation, $\epsilon$ and $\mu$ are respectively the electric and magnetic permeability of the material in the computational domain $\Omega$. $\Omega$ is Lipschitz and piecewise smooth. 

In $\Omega$, the complex time harmonic electric field phasor $\boldsymbol{E}$ satisfies 
\begin{align}\label{eq:phasor}
\nabla \times \mu^{-1} \nabla \times \boldsymbol{E} - \kappa^2 \varepsilon \boldsymbol{E} = 0 \quad \text{in } \Omega, 
\end{align}
where $\kappa = \omega \sqrt{\varepsilon_0 \mu_0}$ is the wave number. 

Additionally, we assume the following generalized impedance boundary condition on $\partial \Omega$, which is expressed as,
\begin{align}\label{eq:bc}
\bm{\hat{n}} \times \mu^{-1} \nabla \times \boldsymbol{E} - i \kappa Z \boldsymbol{E}_T &= Q \big(\bm{\hat{n}} \times \mu^{-1} \nabla \times \boldsymbol{E} + i \kappa Z \boldsymbol{E}_T\big) + \boldsymbol{g} \quad \text{on } \, \partial \Omega,
\end{align}
where, { $\bm{\hat{n}}$ is the normal unit vector  to the boundary surface.} $\boldsymbol{E}_T = \bm{\hat{n}} \times (\boldsymbol{E} \times \bm{\hat{n}})$ is the tangential component of the electric field, $Q$ is a complex scalar-valued function on the boundary with $\lvert Q \rvert \leq 1$, and $\boldsymbol{g}$ is a prescribed source function. The $Z (\ge 0)$ is defined on the boundary $\partial \Omega$, and we take it to be $Z = \sqrt{\lvert \mu \rvert / \lvert \varepsilon \rvert}$. Note that the choice $Q = 1$ corresponds to a Perfect Electric Conductor (PEC) type boundary condition for the waves scattered of the PEC surface and recovered as
\begin{align}\label{eq:PEC_boundarycondition}
\boldsymbol{E}_T = -\frac{1}{2 i \kappa Z} \boldsymbol{g}.
\end{align}

For scattering applications, we assume that we have a bounded scatterer which can be impenetrable (for example, with an impedance or PEC boundary condition) or penetrable (for example, a dielectric) occupying a domain $D$ in the interior of the computational domain $\Omega$. A known incident field $\boldsymbol{E}^i$ impinges on the scatterer and creates an outgoing scattered field $\boldsymbol{E}^s$. Then, the total field $\boldsymbol{E} = \boldsymbol{E}^i + \boldsymbol{E}^s$ satisfies the homogeneous Maxwell system \textcolor{black}{\eqref{eq:Maxwell3DEq}}. The incident field $\boldsymbol{E}^i$ is assumed to satisfy the background Maxwell's equations in the neighborhood of $D$, which is expressed as 
\begin{align}\label{eq:incident_field}
\nabla \times \nabla \times \boldsymbol{E}^i - \kappa^2 \boldsymbol{E}^i = 0.
\end{align}

In this work, we consider the incident plane wave $\boldsymbol{E}^i = \boldsymbol{E}_0^i \exp(i \kappa \boldsymbol{d} \cdot \boldsymbol{x})$, where the real direction vector $\boldsymbol{d}$ satisfies $\|\boldsymbol{d}\| = 1$ and the real polarization $\boldsymbol{E}_0^i \neq 0$ is such that $\boldsymbol{E}_0^i \cdot \boldsymbol{d} = 0$. 

The direct scattering problem involves determining $\boldsymbol{E}^s$ based on the known incident field $\boldsymbol{E}^i$. Traditionally, this requires complex numerical simulations to solve the governing equations. Since these equations depend on the angular frequency and the incident fields can vary with frequency, computing the scattered field for each case using conventional methods is very computationally expensive. To resolve the frequency, we will require multiple grid points for the smaller wavelength. That requires solving a linear system of equation of size $N^3$ where $N$ defines the number of degrees of freedom in the simulations. To address this, we developed a complex-valued DeepONet that efficiently learns the general operator mapping from both frequency and incident field to the total field, which can significantly improve computational efficiency for scattering problems in industrial applications. 


\section{Surrogate Modeling}\label{sec:Model}

In this section, we formulate the neural operators, specifically the DeepONet as a surrogate model to solve the scattering problem described in \autoref{sec:equation_setup}. We first formulate the setup for a vanilla real-valued DeepONet and then design a complex-valued variant to more effectively address the time-harmonic Maxwell's equations. Furthermore, we propose methods to enforce the boundary conditions for the total field, ensuring that the resulting Physics-Informed Complex-Valued DeepONet satisfies these constraints, thus improving the interpretability of the model.

\subsection{ Formulation of a real valued DeepONet}\label{sec:vanilla_DoN}
The DeepONet \cite{Lu_DeepONet} is a neural operator used to learn the underlying mapping between spaces of function from data, based on the universal approximation theorem of operators \cite{chen_universal_thm_neural_operator}. Specifically, we denote the input function by $u: x \mapsto u(x)$, defined on the domain $D \subset \mathbb{R}^n$, and the output function by $v: y \mapsto v(y)$, defined on domain $\Omega \subset \mathbb{R}^m$. Let $\mathcal{U}$ and $\mathcal{V}$ be the function spaces that contain $u$ and $v$, respectively. The mapping from the input function $u$ to the output function $v$ is denoted by an operator:
\begin{align}\label{eq:operator_G}
    \mathcal{G}: \mathcal{U} \ni u \mapsto v \in \mathcal{V}.
\end{align}

The basic DeepONet consists of two primary building blocks: the \textbf{trunk network}, which takes the coordinates $y \in \Omega$ as input and outputs the corresponding basis functions; and the \textbf{branch network}, which takes the discretized input function $u$ as input. Here, $u$ is evaluated at $m$ arbitrary spatial coordinates $\{x_1, x_2, \ldots, x_m \}$ to obtain pointwise evaluations $\boldsymbol{u} = \{u(x_1), u(x_2), \ldots, u(x_m) \}$ that will serve as the input to the branch network. The continuous approximation of the output performed by  DeepONet reads as
\begin{align}\label{eq:DoN_output}
    v(y) = \mathcal{G}(u)(y) \approx \sum_{i = 1}^r b_i(\boldsymbol{u}) t_i(y) + b_0,
\end{align}
where $b_i$ and $t_i$ are the $r$-dimensional outputs of the branch and trunk networks, respectively. $b_0 \in \mathbb{R}$ is a trainable bias term. By evaluating the network's output at the coordinates $\{y_1, y_2, \ldots, y_q \}$, we obtain the continuous prediction of $\bm{v}$ projected at $\bm{y}$ as $[v(y_1), v(y_2), \ldots, v(y_q)]^\top$, which \textcolor{black}{approximates} the ground truth data.

The solutions to the time-harmonic Maxwell's equations are parameterized by the angular frequency. Therefore, our goal is to learn the operator with two inputs $\mathcal{G}: (\boldsymbol{E}^i; f) \mapsto \boldsymbol{E}$, where $\boldsymbol{E}^i \in \mathbb{C}^3$ is the incident field, $f \in \mathbb{R}$ is the angular frequency, and $\boldsymbol{E} \in \mathbb{C}^3$ is the total field. Two branch networks, $b^{inc}$ and $b^{freq}$, are used to encode the incident field and the frequency separately. Additionally, since the incident field is complex-valued, its real and imaginary parts are learned separately in the real-valued DeepONet. In particular, we consider real and imaginary component of a complex valued input and output pair as two separate features and labels, respectively. The total field evaluated at $y \in \mathbb{R}^3$ is given as $\boldsymbol{E}(y) = \mathcal{G}(\boldsymbol{E}^i; f)(y) \approx \sum_{i = 1}^r b^{inc}_i(\boldsymbol{E}^i) b^{freq}_i(f) t_i(y) + b_0$.

\subsection{Formulation of a Complex-valued DeepONet}\label{sec:complex-valued_DoN}
In a real-valued neural network, a complex number $z \in \mathbb{C}$ is represented by 2 real-valued channels, each channel containing the real and imaginary components, $\Re(z)$ and $\Im(z)$ respectively. For example, to learn the mapping $x+iy \mapsto a + ib$ with $N$ samples of data, the real-value neural network learns the mapping between $N$ real-valued 2D vectors
\begin{align}
\begin{bmatrix}
x_1 & y_1 \\
x_2 & y_2 \\
\vdots & \vdots \\
x_N & y_N
\end{bmatrix}
\mapsto
\begin{bmatrix}
a_1 & b_1 \\
a_2 & b_2 \\
\vdots & \vdots \\
a_N & b_N
\end{bmatrix},
\end{align}
and then recombines the real part $a_j$ and imaginary part $b_j$ to produce the complex-valued output $a_j+ib_j$. \textcolor{black}{This separate representation may not capture the phase and magnitude relationships as effectively as treating the complex number as a unified entity (see \autoref{appendix:cvnn_comparison} for an illustrative example)}. In contrast, a complex-valued neural network directly learns the complex-valued mapping
\begin{align*}
    \begin{bmatrix}
x_1 + i y_1 \\
x_2 + i y_2 \\
\vdots  \\
x_N + i y_N
\end{bmatrix}
\mapsto
\begin{bmatrix}
a_1 + i b_1 \\
a_2 + i b_2 \\
\vdots  \\
a_N + i b_N
\end{bmatrix}.
\end{align*}

Previous research \cite{lowe2022complex, virtue2017better, amin2011complex, nitta1997extension} has shown that the real-valued approach to inherently complex-valued data performs poorly in terms of efficient architecture, convergence, and generalization ability. Since the solutions to the time-harmonic Maxwell's equations are inherently complex-valued, we address this challenge by implementing the DeepONet in the complex domain to learn the mapping between complex-valued fields. The complex-valued DeepONet retains the same structure as the basic DeepONet, but both the input and output functions are complex-valued, and all calculations are performed using the \texttt{complex64} datatype.

In a complex-valued neural network (CVNN), the forward pass for a given layer is represented as
\begin{align}\label{eq:forward_pass}
    \boldsymbol{h} & = \boldsymbol{W}\boldsymbol{x} + \boldsymbol{b}, \nonumber \\
    \boldsymbol{z} & = \sigma(\boldsymbol{h})
\end{align}
where $\boldsymbol{x} \in \mathbb{C}^n$ is the input vector, $\boldsymbol{W} \in \mathbb{C}^{m \times n}$ is the complex weight matrix, $\boldsymbol{b} \in \mathbb{C}^m$ is the complex bias vector, and $\sigma(\cdot)$ is the activation function. 

Two types of activation functions exist for CVNN. First, split activation functions, where real and imaginary parts are treated independently; second, fully complex activation function, where real and imaginary parts are treated as a single entity. A complex-valued activation function cannot be bounded and analytic everywhere at the same time, because Liouville's theorem states that such functions are constant. Due to the computation complexity and the difficulty in fulfilling Liouville's theorem, fully complex activation functions are less frequently studied than split-type activation functions \cite{Lee_CVNN_Survey}. 

In complex-valued DeepONet, we adopt a split ReLu activation function, denoted $\mathbb{C}\text{ReLu}$, that applies $\text{ReLu}$ to the real and imaginary parts of the complex-valued input $z \in \mathbb{C}$
\begin{align}\label{eq:activation_fn}
    \mathbb{C}\text{ReLu}(z) = \max(0, \Re(z)) + i \cdot \max(0, \Im(z)).
\end{align}
The $\mathbb{C}\text{ReLu}$ function satisfies the Cauchy-Riemann equations when both the real and imaginary parts are at the same time either strictly positive or strictly negative but is not bounded. 

Additionally, to enhance DeepONet's ability to capture oscillatory patterns in the functions, a periodic feature-extension layer is added to the trunk network input $\boldsymbol{y}$. This layer computes multiple sine and cosine values of the input before passing the augmented features to the rest of the trunk network. The periodic feature extension is expresses as
\begin{align}\label{eq:feature_expansion}
    \text{periodic feature of } \boldsymbol{y} = \boldsymbol{y} + \sum_{n=1}^{N} \sin(n\boldsymbol{y}) + \cos(n\boldsymbol{y}),
\end{align}
where $N$ is the extended feature size. 

Finally, a real-valued loss function for the complex-valued DeepONet is designed to account for both the real and imaginary parts of the output. As an example, let $E_x^{NN}(x,y,z)$ denote the DeepONet prediction of the $x$-component of the incident electric field, evaluated at the coordinate $(x,y,z)$, and let $\bar{E}_x(x,y,z)$ denote ground truth total field. The mean squared error (MSE) for the complex data is defined as follows
\begin{align}\label{eq:MSE_loss}
    \text{MSE}(E_x^{NN}, \bar{E}_x) = \frac{1}{N} \sum_{(x,y,z) \in \Omega} \left[ (\Re(E_x^{NN}(x,y,z)) - \Re(\bar{E}_x(x,y,z)))^2 + (\Im(E_x^{NN}(x,y,z)) - \Im(\bar{E}_x(x,y,z)))^2 \right],
\end{align}
where $N$ is the total number of data points in the computational domain $\Omega$. 

The same MSE function is used for all other components of the electric and magnetic fields. A separate DeepONet is implemented for each of the six components of the electric and magnetic fields to minimize their losses separately. Our implementation utilizes the Complex Valued Neural Network package \cite{cvnn_Zenodo_package} to simplify the workflow. { To illustrate the advantages of the complex-valued DeepONet over the real-valued counterpart, we present a detailed analysis along with computational results for a representative complex-domain function approximation. The complete comparison and supporting results are provided in \autoref{appendix:cvnn_comparison}, highlighting the improved accuracy and stability achieved by the complex-valued architecture.}

\subsection{Imposing the PEC boundary condition}\label{sec:boundary_condition}
To impose the physics on the proposed DeepONet architecture, we adopt the approach of soft constraint. \textcolor{black}{Since the metallic scatterer examples studied in this work are perfect electric conductors (PEC), the electric field on the surface of a PEC scatterer is orthogonal to the conductor.} Therefore, the PEC boundary condition entails that $\bm{\hat{n}} \times \bm{E} = 0$, where $\bm{\hat{n}}$ is the normal vector to the scatterer surface and $\boldsymbol{E}$ is the total field. Previous research such as \cite{ling2016machine, ling2016reynolds} has proposed ways of integrating such first-principle-informed conditions directly into the architecture of neural networks. Due to the high-dimensional training data in our problem, we instead incorporate the PEC boundary condition directly into the loss function.

\begin{figure}[h!]
    \centering
    \includegraphics[width=0.7\linewidth]{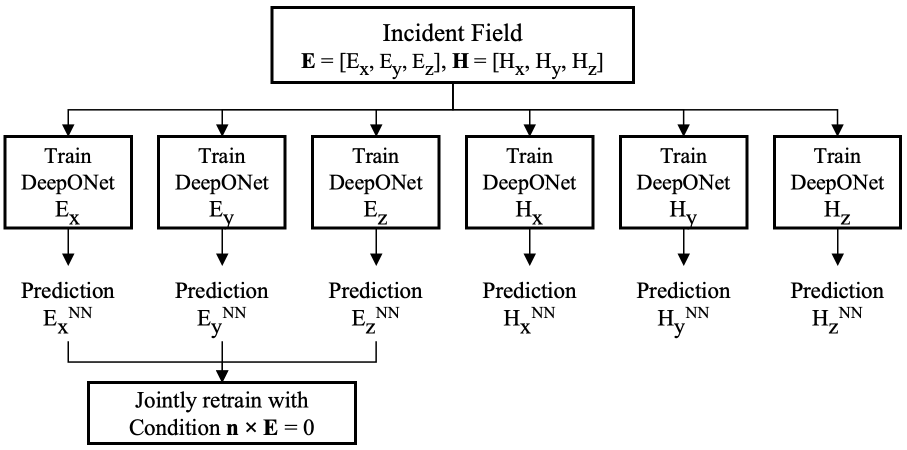}
    \caption{Illustration of the two-stage process for training a combination of DeepONet models for different field components to impose the PEC boundary condition $\bm{\hat{n}} \times \bm{E} = 0$ on the electric field.}
    \label{fig:nxE_diagram}
\end{figure}

Since a separate complex-valued DeepONet is trained for each field component, the boundary condition is enforced in two stages, as illustrated in \autoref{fig:nxE_diagram}. First, DeepONets are trained independently and concurrently for the components $E_x$, $E_y$, $E_z$, $H_x$, $H_y$, and $H_z$, with each model minimizing the data-driven loss as defined in \textcolor{black}{\eqref{eq:MSE_loss}}. Once the data-driven loss for each component is minimized sufficiently, the trained models are saved. Next, to impose the boundary condition on the electric field, previously trained models for $E_x$, $E_y$, and $E_z$ are loaded and retrained jointly. During this  phase of training, the PEC boundary condition is introduced into the loss function as an additional term. The loss function for this retraining step is defined as
\begin{align}\label{eq:PEC_loss}
    L = w_1 || \bm{\hat{n}} \times \bm{E}^{NN} ||_2^2 + w_2 \left( \text{MSE}(E_x^{NN}, \bar{E}_x) + \text{MSE}(E_y^{NN}, \bar{E}_y) + \text{MSE}(E_z^{NN}, \bar{E}_z) \right),
\end{align}
where $|| \cdot ||_2$ represents the $L_2$ norm. The weights $w_1$ and $w_2$ control the balance between the boundary-condition-informed and data-driven losses and are adjusted during retraining to ensure both the boundary condition and the data loss are satisfied.

\subsection{Reparametrization of the Complex-valued DeepONet}\label{sec:two_step}

In a complex-valued DeepONet, the trunk network serves as the basis functions for the output, while the branch network provides the coefficients. In \cite{Shin_two_step_DppONet}, the study introduced a two-step training method to simplify the optimization of the DeepONet: the first step is to determine the basis representation through the trunk network and find the corresponding coefficients without involving the branch network; and in the second step, the branch network learns these coefficients. We can extend this method into the complex domain, as the operations are well-defined for complex matrices. 

Specifically, let $\bm{u} = \{u_k\}_{k=1}^K$ be a set of complex-valued input functions from the function space $\mathcal{U}$, and let $v_k(\cdot) = \mathcal{G}[u_k](\cdot)$ be the corresponding complex-valued output functions in the space $\mathcal{V}$. The branch network, denoted as $\bm{b}(\cdot; \theta)$, is a CVNN with $L_b$ layers and an output dimension of $N$
\begin{align}\label{eq:branch}
    \bm{b}(\cdot; \theta) = (b_0(\cdot; \theta), \ldots, b_N(\cdot; \theta))^\top \in \mathbb{C}^{N},
\end{align}
where $\theta$ represents the parameters of the branch network. Similarly, the trunk network, denoted as $\bm{t}(y; \mu)$, is a vector-valued CVNN with $L_t$ layers defined for $y \in \Omega_y \subset \mathbb{C}^{d_y}$ with the same output dimension
\begin{align}\label{eq:trunk}
    \bm{t}(y; \mu) = (1, t_1(y; \mu), \ldots, t_{N-1}(y; \mu))^\top \in \mathbb{C}^{N},
\end{align}
where $\mu$ represents the parameters of the trunk network.

The output of the complex-valued DeepONet is given by the element-wise product of the branch and trunk networks:
\begin{align}\label{eq:output}
    O_{\text{net}}[\bm{u}, \Theta] = \bm{t}^\top(y; \mu) \bm{b}(\bm{u}; \theta),
\end{align}
where $\Theta = \{\theta, \mu\}$ is the set of all trainable parameters in the DeepONet. When training the DeepONet with a batch of input functions $\{u_k\}_{k=1}^K$ and trunk net coordinates $\{y_i\}_{i=1}^{m}$, we denote
\begin{align}\label{eq:trunk_matrix}
    \bm{T}(\mu) := \begin{bmatrix}
        \bm{t}^\top(y_1; \mu)\\
        \vdots \\
        \bm{t}^\top(y_m; \mu)
    \end{bmatrix} \in \mathbb{C}^{m \times N},
\end{align}
\begin{align}\label{eq:branch_matrix}
    \bm{B}(\theta) := [\bm{b}(u_1; \theta), \ldots, \bm{b}(u_K; \theta)] \in \mathbb{C}^{N \times K},
\end{align}
and the matrix of target outputs $\bm{V} := [\bm{v}_1, \ldots, \bm{v}_K] \in \mathbb{C}^{m \times K}$. 

In the first step of the two-step method, the trunk network is trained by solving the following optimization problem:
\begin{align}
    \min_{\mu, A} \mathcal{L}(\mu, A) := \left\|\bm{T}(\mu)A - \bm{V}\right\|_2^2,
\end{align}
where $A \in \mathbb{C}^{N \times K}$ is a complex matrix that temporarily replaces the branch net during the phase. Let $(\mu^*, A^*)$ be an optimal solution to this problem and let $\bm{T}(\mu^*)$ be full rank. We can perform a QR-factorization of the optimized trunk network, i.e., $Q R = \bm{T}(\mu^*)$, where $Q \in \mathbb{C}^{m \times m}$ is a unitary matrix (the conjugate transpose $Q^\dagger = Q^{-1}$) and $R \in \mathbb{C}^{m \times N}$ is an upper triangular matrix. 

In the second step, we train the branch network to fit the product $RA^*$. This is done by solving the optimization problem:
\begin{align}
    \min_{\theta} \left\|\bm{B}(\theta) - RA^*\right\|_2^2.
\end{align}
Assuming $\theta^*$ to be the optimal solution, the fully trained branch network is then given by $\bm{b}(\cdot; \theta^*)$. 

\subsection{Sampling of the input space}\label{sec:acq_function}

The performance of the DeepONet is sensitive to the values of the angular frequency selected for training. However, sampling a broad range of frequencies is inefficient and impractical due to the computational expense of generating large training datasets. Thus, an efficient data acquisition strategy is required to select frequencies that maximize information gain and enhance model performance with fewer samples. Traditional methods for sampling training frequencies within a given interval $[f_{min}, f_{max}]$ such as uniform sampling, Latin hypercube sampling, and random sampling do not account for the sensitivity of the model's predictions to input frequencies. Instead, we employ the method of acquisition function \cite{Acquisition_func_pickering_paper} to sample input space for the training of complex-valued DeepONet. 

An acquisition function is a criterion used in active learning to select the next data point to be sampled. The motivation behind using an acquisition function is to efficiently explore the input space, especially in scenarios involving rare and extreme events, and construct a more generalizable model with fewer samples. To implement the acquisition function, we employ an ensemble of $N$ DeepONets, where the mean solution of the ensemble, $\bar{G}(f)$, with respect to input frequency $f$ is given by:
$$
\bar{G}(f) = \frac{1}{N} \sum_{n=1}^{N} G_n(f),
$$
where $G_n(f)$ denotes the prediction of the $n$-th DeepONet in the ensemble. The predictive variance is then calculated as
\begin{align}
    \sigma^2(f) = \frac{1}{N-1} \sum_{n=1}^{N} |G_n(f) - \bar{G}(f)|^2.
\end{align}

which gives a real scalar representing the spread of the complex-valued predictions. We then define the acquisition function to be this predictive variance
\begin{align}
a(f) = \sigma^2(f),
\end{align}
During the training, Monte Carlo optimization is employed to find the samples where the predictive variance is the greatest. This is done by generating a set of random samples from the input space using Latin Hypercube Sampler and evaluating them with the DeepONet ensemble to identify the sample that maximizes the acquisition function. The identified sample is then incorporated into the training dataset. This method prioritizes sampling points with the highest predictive variance, so that the training focuses on regions of the input space where the model is most uncertain and provides the greatest gain for learning about the underlying system. 

\section{Computational Experiments} \label{Results}
{
In this section, we demonstrate the performance of the proposed DeepONet architecture in solving Maxwell's equations. To evaluate its effectiveness across different geometric complexities, we consider two distinct geometries: 
\begin{enumerate}
    \item A three-dimensional sphere with a smooth surface, which serves as a baseline case without singularities.
    \item An almond-shaped geometry with sharp corners, which introduces singularities and poses a greater challenge for the neural network to accurately approximate the electromagnetic fields.
\end{enumerate}
This comparison allows us to assess the capability of the proposed DeepONet to handle both smooth and singularity-prone geometries.
}
\subsection{3D Homogeneous Metallic Sphere }\label{sec:DoN_results}
Here, we investigate the performance of the proposed complex-valued DeepONet architecture and highlight its advantages over conventional real-valued DeepONets. Specifically, we compare the two approaches in terms of accuracy, convergence, and their ability to capture complex-valued solutions arising from electromagnetic problems at two very large frequencies. This analysis demonstrates the superiority of the complex-valued framework, particularly in handling the inherent coupling between the real and imaginary components of the solution, which is often lost when using real-valued networks. 

\begin{figure}[h!]
    \centering
    \begin{subfigure}[b]{0.45\textwidth}
        \includegraphics[scale=0.4]{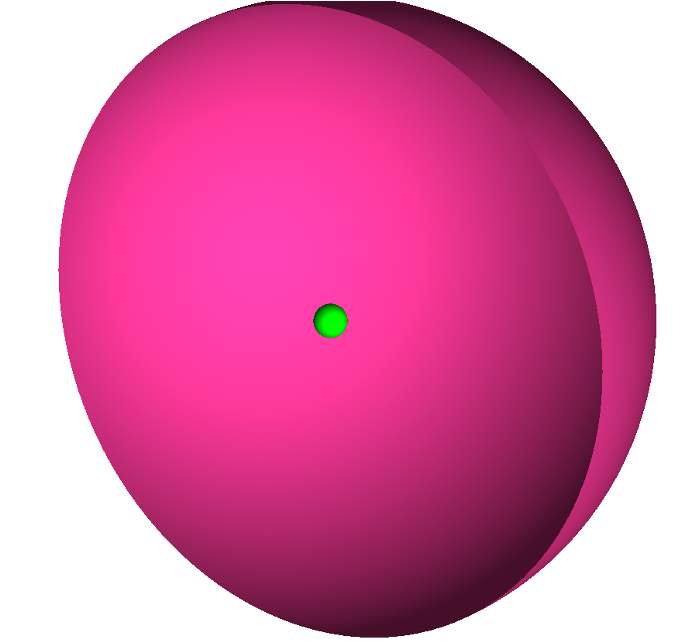}
        \caption{Computational domain with the outer boundary and metallic sphere.}
        \label{fig:geometry1}
    \end{subfigure}
    \hfill
    \begin{subfigure}[b]{0.45\textwidth}
        \includegraphics[scale=0.4 ]{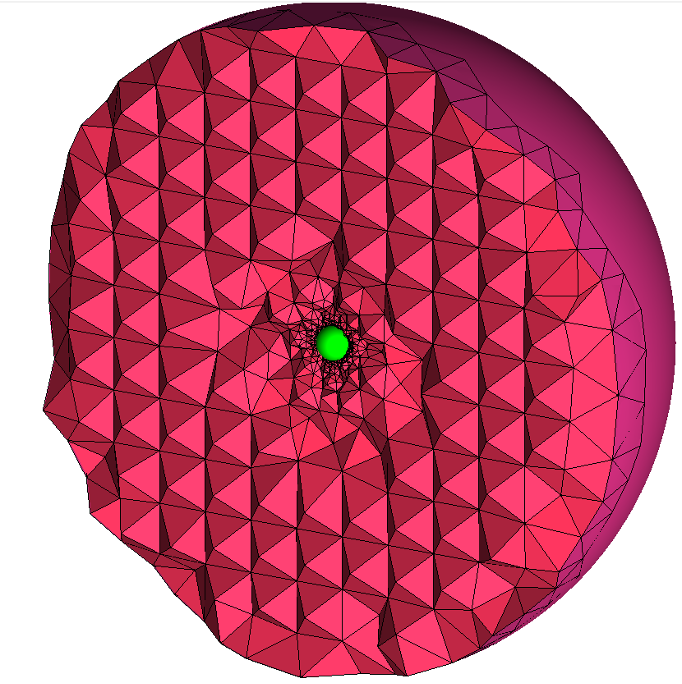}
        \caption{Meshed geometry of the computational domain.}
        \label{fig:geometry2}
    \end{subfigure}
    \caption{Computational domain used for electromagnetic simulations. Subfigure (a) shows the simplified outer boundary with a metallic sphere (in green color) at the center. Subfigure (b) shows  crinkled view of the tetrahedral meshes used to discretize the computational domain show in (a).}
    \label{fig:geometry}
\end{figure}

The computational domain of the problem, shown in \autoref{fig:geometry}, consists of a homogeneous metallic sphere with a radius of $r = 1 \, \text{m}$ (shown as green sphere) and an outer boundary positioned at a distance of $3\lambda$ from the metallic sphere, where $\lambda  \, \text{(m)}$ is the wavelength given as 
\begin{align}
    \lambda  = \frac{c}{f}
\end{align}
where $c = 3 \times 10^8 \, \text{m/s}$ is the speed of light and $f \, \text{(Hz)}$ is the maximum frequency of the electromagnetic wave considered in this problem. 

The computational domain is discretized by tetrahedral meshes with $3$ cells per wavelength $(\lambda)$ and is presented in \autoref{fig:geometry2}. Two different problems are considered in this work. For the first study, we consider electromagnetic fields with a maximum frequency of $0.6 \, \text{GHz}$. The corresponding wave length $\lambda = 0.5~\text{m}$ and the sphere surface resolution is $\frac{1}{6} \, \text{m}$. Under this configuration, the sphere surface is discretized into 1004 triangular facets, the outer boundary is discretized into 1408 triangular facets, and the volume cells are discretized into 25,042 tetrahedral elements with a maximum cell size of $6\lambda$. For a more advanced second study, we consider a much higher maximum frequency of $4.77 \, \text{GHz}$, so the corresponding wavelength $\lambda = 6.28 \times 10^{-2} \, \text{m}$, the sphere surface resolution is $2.10 \times 10^{-2} \, \text{m}$, and the sphere is discretized into 137,388 triangular facets. In the second case, we test the scalability of the proposed surrogate model to a significantly larger dataset required to resolve the higher frequency and smaller wavelength.

The training dataset contains the incident fields $\boldsymbol{E}^i(x,y,z;f)$ impinging the sphere at different incidence angles, frequencies $(f)$, and spatial locations of sensors, which in this case are coordinates of vertices $(x, y, z)$ of the tetrahedral and triangles. The output of the DeepONet is the total field $\boldsymbol{E}(x,y,z; f)$. To generate the training and testing data, we utilized the Ultra Weak Variational Formulation (UWVF) method \cite{Huttunen_2002,Huttunen_2007,Huttunen_2018} to solve the system described by \textcolor{black}{\eqref{eq:Maxwell3DEq}}. UWVF is a special case of a Trefftz-discontinuous Galerkin (Trefftz-DG) method provided the scattering medium is lossless. The UWVF approach uses an unstructured finite element computational grid and a superposition of plane wave solutions of Maxwell's equations on each tetrahedral element \cite{Luostari_2013}. UWVF has been found to be efficient for approximating solutions of Maxwell’s equations for a wide range of test problems as shown by Lahivaara et al.\cite{lahivaara2024high}. 

\begin{figure}[h!]
    \centering
    \begin{subfigure}[t]{0.48\textwidth}
        \centering
        \includegraphics[width=\textwidth]{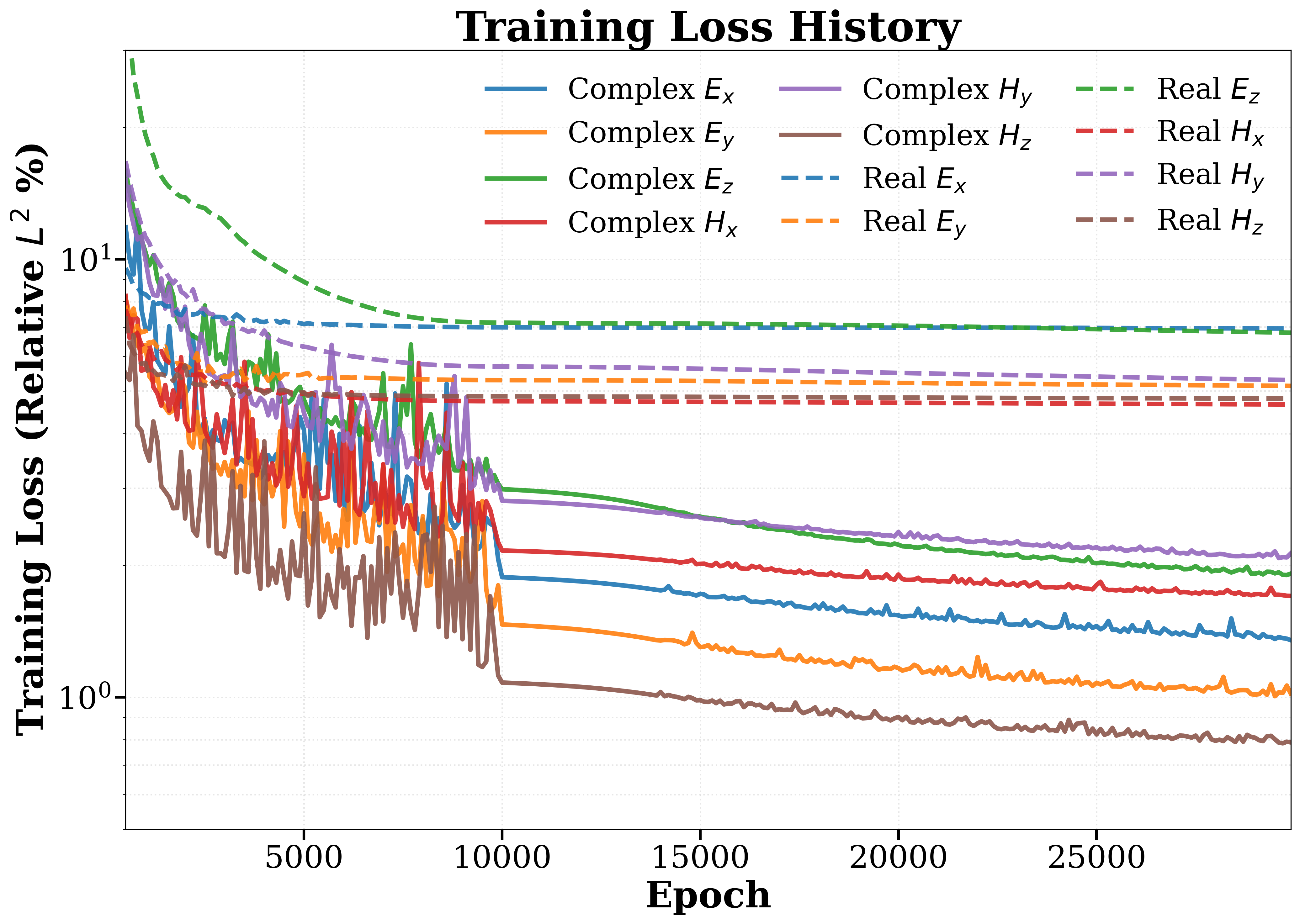}
        \label{fig:loss_history}
    \end{subfigure}
    \hspace{5pt}
    \begin{subfigure}[t]{0.48\textwidth}
        \centering
        \includegraphics[width=\textwidth]{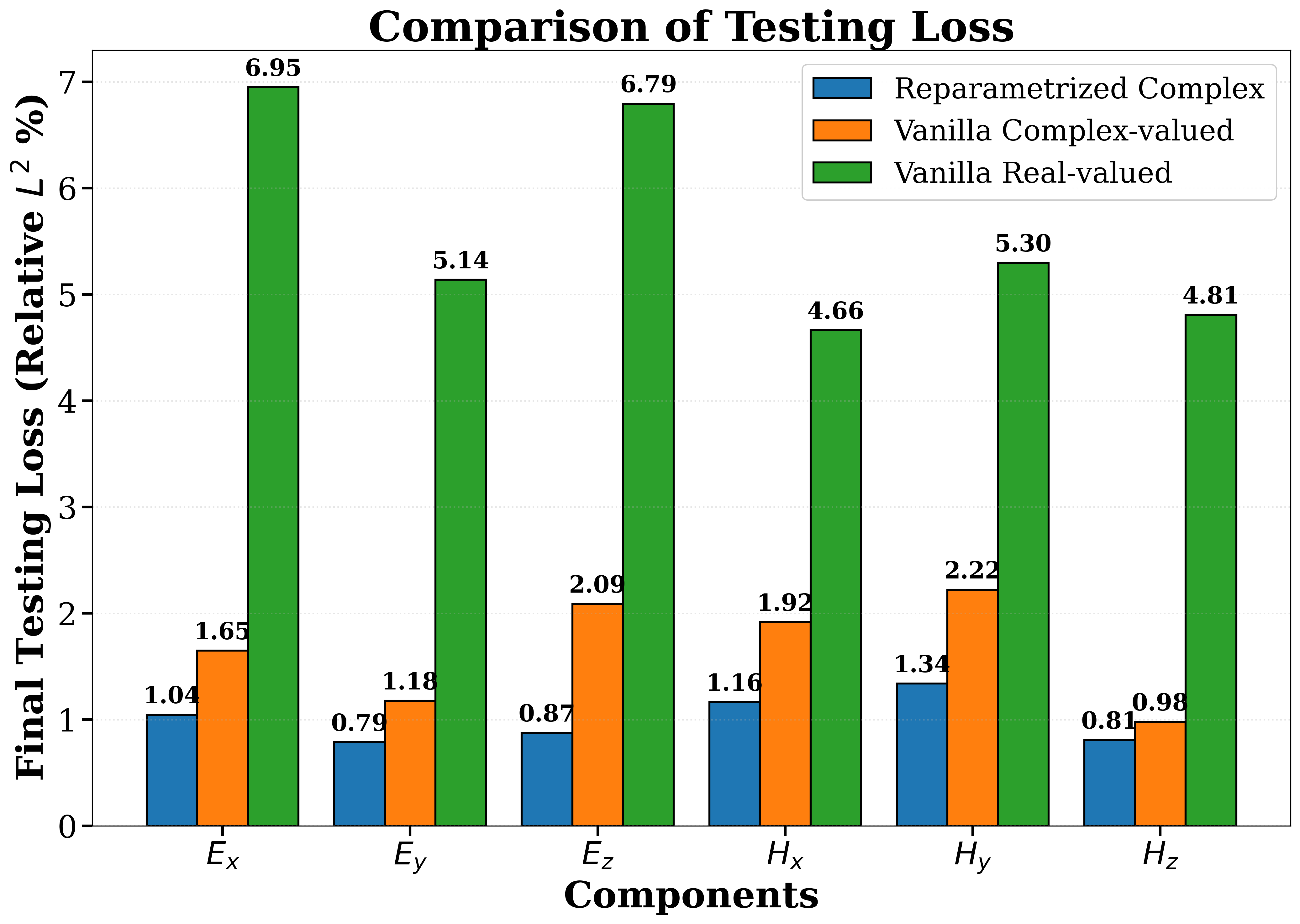}
        \label{fig:compare_loss}
    \end{subfigure}
    \caption{{Left subfigure shows training loss history for each electromagnetic field using complex-valued DeepONets (solid lines) and real-valued DeepONets (dashed lines), starting from epoch 500 on the logarithmic scale. Each field component is distinguished by color. The figure shows the superior convergence of the complex-valued DeepONet. The right subfigure shows testing performance of DeepONets (the complex-valued DeepONet and the reparametrized complex-valued DeepONet). The plot shows a superior performance of complex-valued DeepONets over real-valued DeepONet.}}
    \label{fig:performance_comparison}
\end{figure}

To demonstrate the superior performance of the complex-valued DeepONet compared to the vanilla real-valued DeepONet, we used similar architectures for both models, training them on the same 80 frequencies and testing on 21 frequencies sampled from a range of $[0.05, 0.6] \, \text{GHz}$. The hyperparameters for both models used in our experiments are summarized in \autoref{tab:hyperparam}. For the real-valued DeepONet, the real and imaginary parts of a function are trained separately and then combined into a complex number for comparison against the ground-truth data. 
{Note that while the complex-valued DeepONet uses complex parameters (effectively doubling the real parameter count per weight), the real-valued DeepONet requires two separate networks—one for the real part and one for the imaginary part—for each field component. Consequently, the total number of trainable real-valued parameters is approximately the same for both architectures.}
The training loss histories of the two models are plotted in \autoref{fig:performance_comparison}, with the dotted line representing the real-valued DeepONet and the solid line representing the complex-valued DeepONet. The loss stabilizes at around 10,000 epochs, with the complex-valued DeepONet achieving superior training loss across all components of the field. The final testing results are summarized in \autoref{fig:performance_comparison}, which shows that the reparametrized complex-valued DeepONet, trained using the two-step method, exhibits the best testing performance. The complex-valued models significantly outperform the vanilla real-valued model by decreasing the error with a factor ranging from 5-7 times. In terms of inference speed, the UWVF numerical solver takes 15 minutes to predict one frequency using 192 cores. In comparison, the complex-valued DeepONet predicts a single frequency in 1.16 seconds on a 24-core CPU or 0.10 seconds on an Nvidia Titan RTX GPU.

\begin{figure}[h!]
    \centering
    \includegraphics[width=0.75\textwidth]{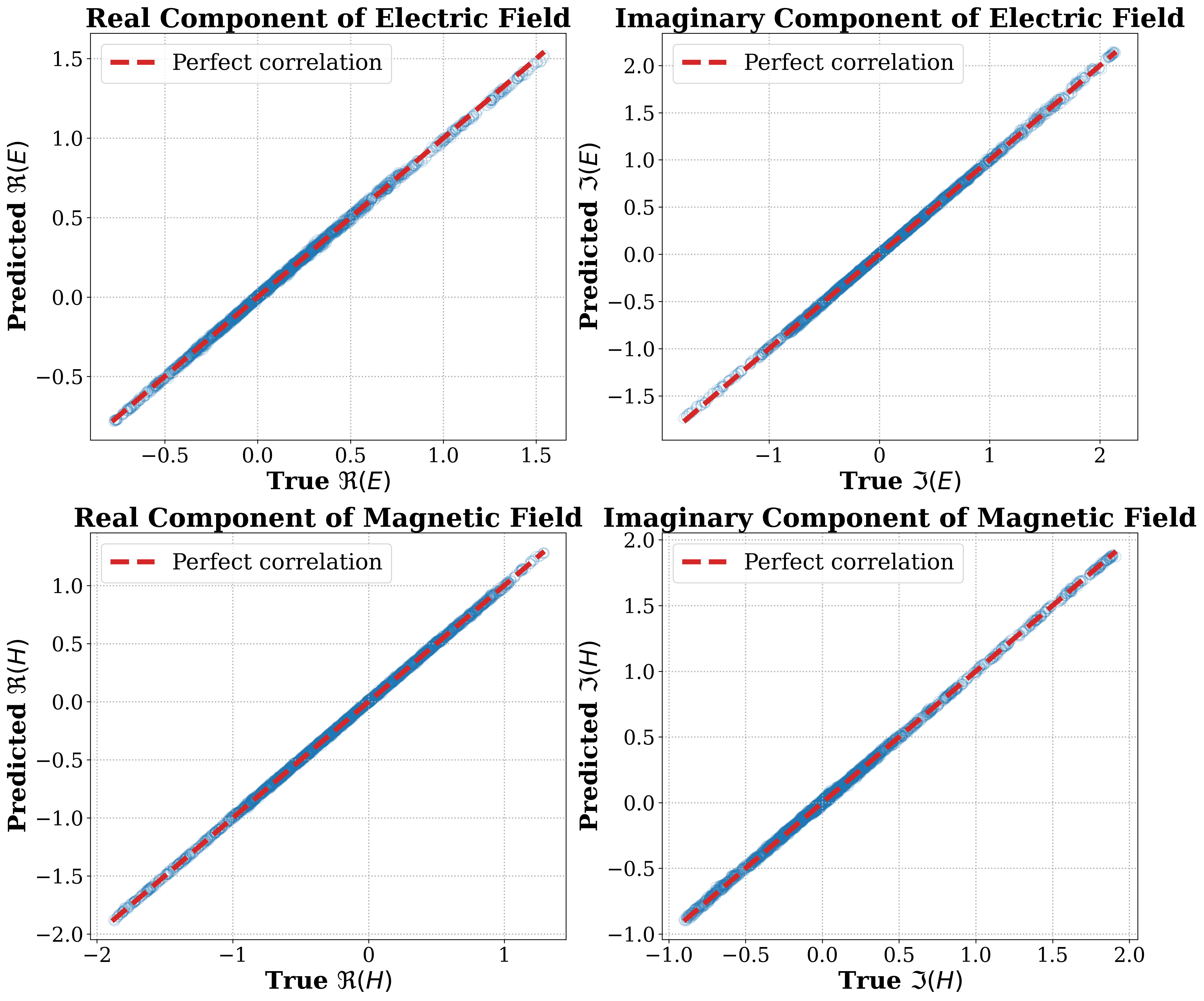}
    \caption{Scatter plots showing the correlation between predicted and true values for the real ($\Re (E)$) and imaginary ($\Im(E)$) components of the electric field, and for the real ($\Re(H)$) and imaginary ($\Im(H)$) components of the magnetic field. Strong diagonal alignment demonstrates high prediction accuracy for both fields.}
    \label{fig:correlation}
\end{figure}

\begin{figure}[!h]
    \centering
    \includegraphics[width=0.95\textwidth]{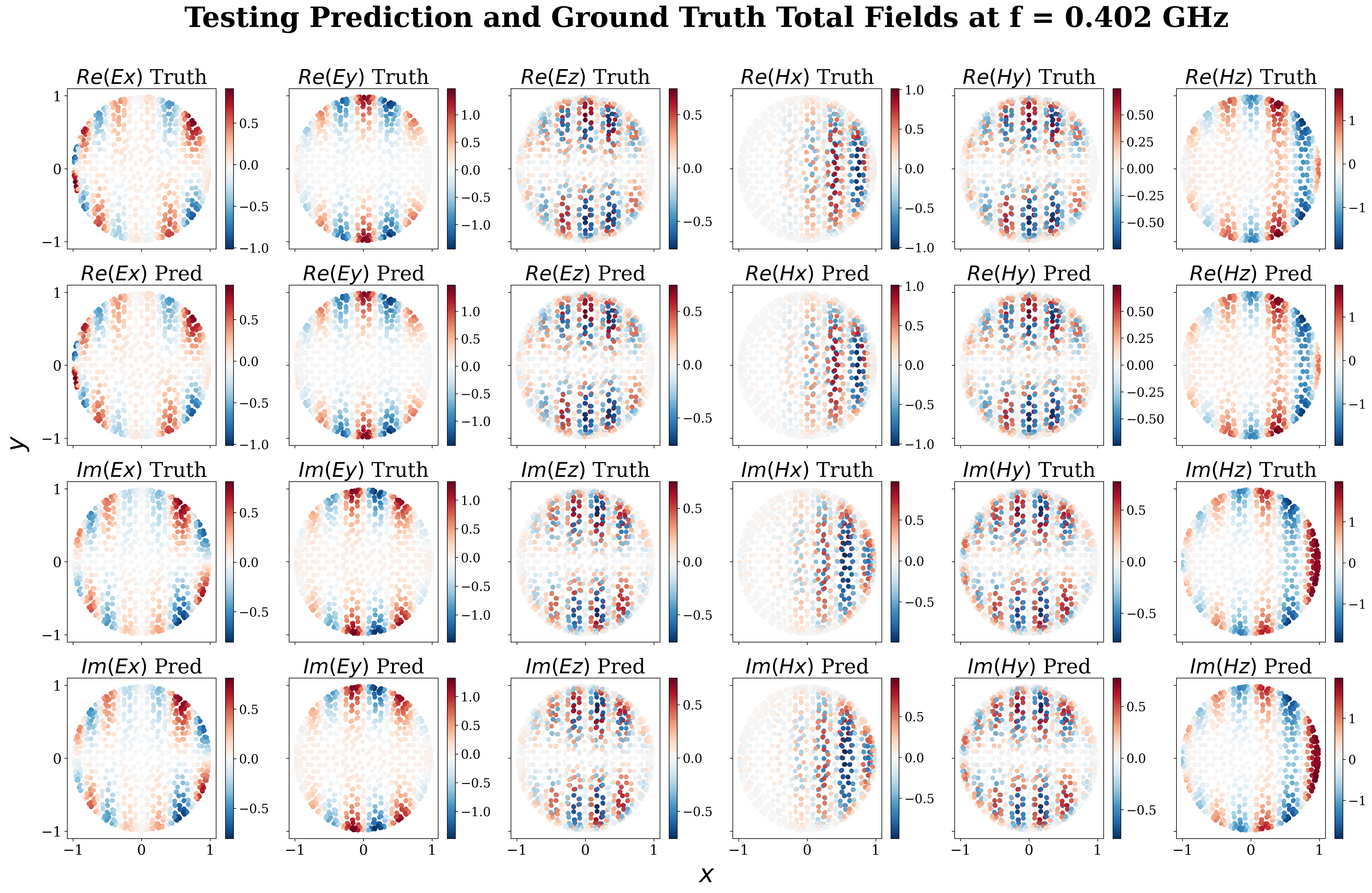}
    \caption{Comparison between ground truth and predictions of the electromagnetic fields on the sphere geometry at a testing frequency of 0.402 GHz. The top four rows show the real and imaginary parts of the electric field components (top two rows) and magnetic field components (bottom two rows). Each column corresponds to a field component (Ex, Ey, Ez, Hx, Hy, Hz). The predictions agree closely with the ground truth across all components.}
    \label{fig:sphere_predictions}
\end{figure}

\begin{figure}[!h]
    \centering
    \includegraphics[width=0.99\textwidth]{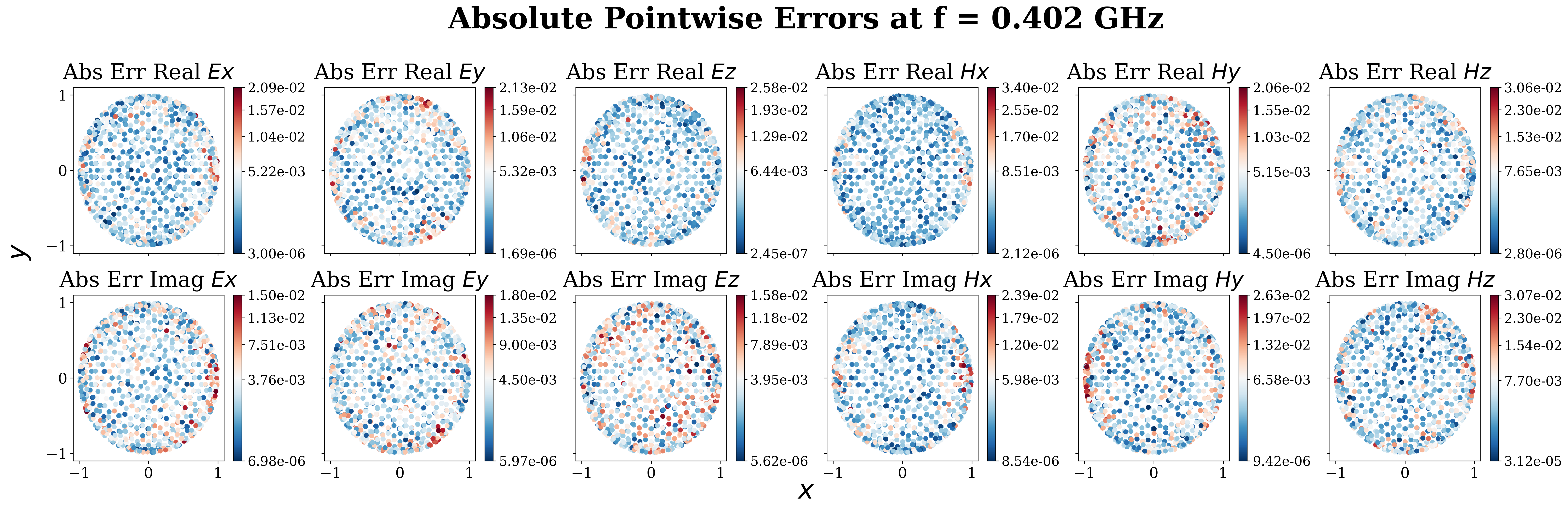}
    \caption{Absolute pointwise errors for all six electromagnetic field components on the sphere geometry at a testing frequency of 0.402 GHz. The top row shows errors in the real parts and the bottom row shows errors in the imaginary parts of each component. The errors are small throughout the domain.}
    \label{fig:sphere_errors}
\end{figure}

\autoref{fig:correlation} shows the correlation between predicted and true total fields. Strong diagonal alignment demonstrates high prediction accuracy for both the electric and magnetic fields. \autoref{fig:sphere_predictions} shows the predictions of the total electric and magnetic fields by the complex-valued model at a testing frequency of 0.402 GHz. The 2D visualizations on the sphere surface demonstrate close agreement between predictions and ground truth for both real and imaginary parts of all field components. \autoref{fig:sphere_errors} presents the absolute pointwise errors, which are small throughout the domain. The model is shown to generalize well to the testing frequencies.

\begin{table}[!h]
\caption{Hyperparameters of the vanilla real-valued and complex-valued DeepONets}
\vspace{1 mm}
\centering
    \begin{minipage}{.5\linewidth}
      \centering        
            \begin{tabular}{|l|l|} \hline 
                 \multicolumn{2}{|c|}{\textbf{Real Valued DeepONet}}\\ \hline 
                 Branch net architecture& $ [1004, 64, 64, 64, 64, 200] $ \\ \hline 
                 Trunk net architecture&  $ [3, 64, 64, 64, 200] $ \\ \hline 
                 Activation& $\text{ReLu}$ \\ \hline 
                 Latent dimension $r$& 200 \\ \hline 
                 Optimizer& Adam \\ \hline 
                 Learning rate& \makecell[l]{Polynomial learning rate \\ schedule, with initial and \\ end learning rate at $10^{-3}$ \\ and $10^{-4}$, respectively} \\ \hline 
                 Training epochs& 30000 \\ \hline
            \end{tabular}
    \end{minipage}%
    \begin{minipage}{.5\linewidth}
      \centering
            \begin{tabular}{|l|l|} \hline 
                 \multicolumn{2}{|c|}{\textbf{Complex Valued DeepONet}}\\ \hline 
                 Branch net architecture& $ [1004, 64, 64, 64, 64, 200] $\\ \hline 
                 Trunk net architecture& $ [3, 64, 64, 64, 200] $\\ \hline 
                 Activation& $\mathbb{C} \text{ReLu}$\\ \hline 
                 Latent dimension $r$ & 200\\ \hline 
                 Optimizer& Adam\\ \hline 
                 Learning rate& \makecell[l]{Polynomial learning rate \\ schedule, with initial and \\ end learning rate at $10^{-3}$ \\ and $10^{-4}$, respectively} \\ \hline 
                 Training epochs& 30000 \\ \hline
            \end{tabular}
    \end{minipage} 
    \label{tab:hyperparam}
\end{table}

In the second example, we tested the capability of the complex-valued DeepONet to handle electromagnetic fields with a much higher maximum frequency of $4.77 \, \text{GHz}$. To resolve the higher frequencies and smaller wavelengths, the computational domain is discretized into a finer mesh as described in \autoref{sec:data_generation} and the incident field is evaluated on 137,388 sensor locations. To accommodate this increase, a larger DeepONet model is created with a branch network input dimension of 137,388 and the number of neurons per layer raised to 128. The number of hidden layers and other model hyperparameters are unchanged. After training for 100,000 epochs, the training and testing relative $L_2$ losses are presented in \autoref{large_dataset_l2}. These results show that the complex-valued DeepONet can effectively scale to a much larger dataset to handle high-frequency fields without compromising its accuracy or altering training routine. 
\begin{table}[h!]
\centering
\caption{Relative $L_2$ errors for training and testing for complex-valued DeepONet on a much larger dataset with high-frequency fields up to $4.77 \, \text{GHz}$} \label{large_dataset_l2}
\vspace{1mm}
\begin{tabular}{|c|c|c|}
\hline
\textbf{Component} & \textbf{Training $L_2$ \%} & \textbf{Testing $L_2$ \% } \\
\hline
$E_x$& 2.94 & 3.01 \\
\hline
$E_y$& 1.34 & 1.33 \\
\hline
$E_z$& 1.99 & 1.97 \\
\hline
$H_x$& 1.78 & 1.78 \\
\hline
$H_y$& 1.15 & 1.15 \\
\hline
$H_z$& 2.74 & 2.71 \\
\hline
\end{tabular}
\end{table}

\textcolor{black}{\subsection{3D Almond shape geometry with sharp corners}\label{sec:almond_results}}

\textcolor{black}{To further demonstrate the capability of the complex-valued DeepONet on geometries beyond the simple sphere, we consider the metallic Almond-shaped target. The Almond target is 0.252 m long and features sharp edges and corners along its boundary, making it more challenging than the smooth spherical scatterer. These sharp features are known to be difficult for numerical methods as they can cause field singularities and require higher mesh resolution to capture accurately.}

\begin{figure}[h!]
    \centering
    \includegraphics[width=0.6\textwidth]{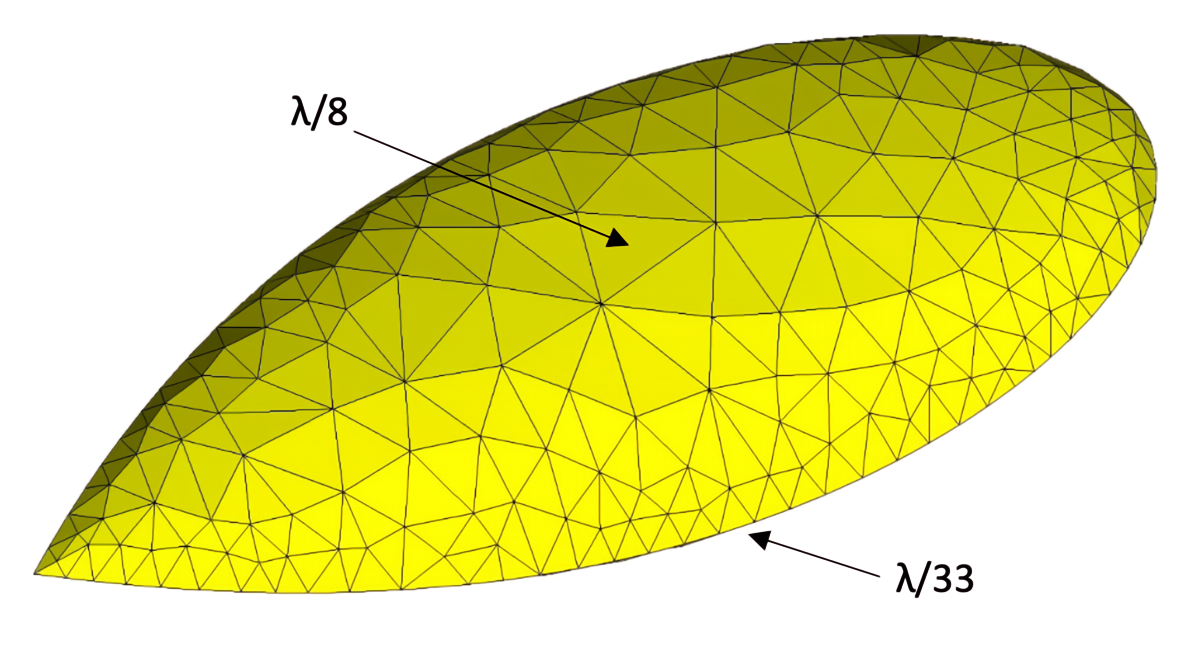}
    \caption{\textcolor{black}{Three-dimensional view of the Almond geometry with an unstructured tetrahedral mesh. The mesh resolution is $\lambda/8$ on the target surface and $\lambda/33$ on the sharp edges to accurately capture the geometry curvature at 1.19 GHz.}}
    \label{fig:almond_geometry}
\end{figure}

\textcolor{black}{The computational domain is generated based on analytical expressions provided by the ElectroMagnetics Code Consortium (EMCC). Following the mesh parameters used for sphere geomtery in in \autoref{sec:DoN_results}, the Almond surface is discretized with a resolution of $\lambda/8$ on the target surface and $\lambda/33$ on the sharp edges. The outer boundary is positioned at $3\lambda$ from the target, where $\lambda$ is the wavelength at the maximum frequency of 1.19 GHz. The incident plane wave impinges on the Almond at an azimuthal angle of $\phi = 60^{\circ}$ in the $\theta = 90^{\circ}$ plane. The computational domain is discretized into 11,154 tetrahedral elements, resulting in the evaluation of the incident field on 11,154 sensor locations. A three-dimensional view of the meshed Almond geometry is shown in \autoref{fig:almond_geometry}.}

\textcolor{black}{The training and testing dataset are generated using the UWVF method for 61 uniformly sampled incident angles. The dataset is split evenly, with 50\% of the data used for training on a uniform grid and 50\% for testing. Six separate complex-valued DeepONets are trained for the six components of the electromagnetic field ($E_x$, $E_y$, $E_z$, $H_x$, $H_y$, $H_z$). The network architecture for each component consists of two branch networks with 4 hidden layers of 256 neurons each, and a trunk network with 3 hidden layers of 256 neurons. The branch networks encode the incident field and frequency information into a 200-dimensional latent space, while the trunk network processes spatial coordinates augmented with Fourier features (feature size = 5). All networks use the complex ReLU ($\mathbb{C}\text{ReLU}$) activation function. The models are trained with a single Adam optimizer using a polynomial learning rate schedule, starting at $10^{-3}$ and decaying to $10^{-5}$ over 10,000 steps with a decay power of 0.5. The training is conducted for up to 200,000 epochs for the $E_x$ component and 100,000 epochs for all other components.}

\textcolor{black}{Note that since the almond geometry is significantly more difficult to predict than the sphere geometry, we employ a larger network and train for more epochs. }

\begin{figure}[h!]
    \centering
    \includegraphics[width=0.9\textwidth]{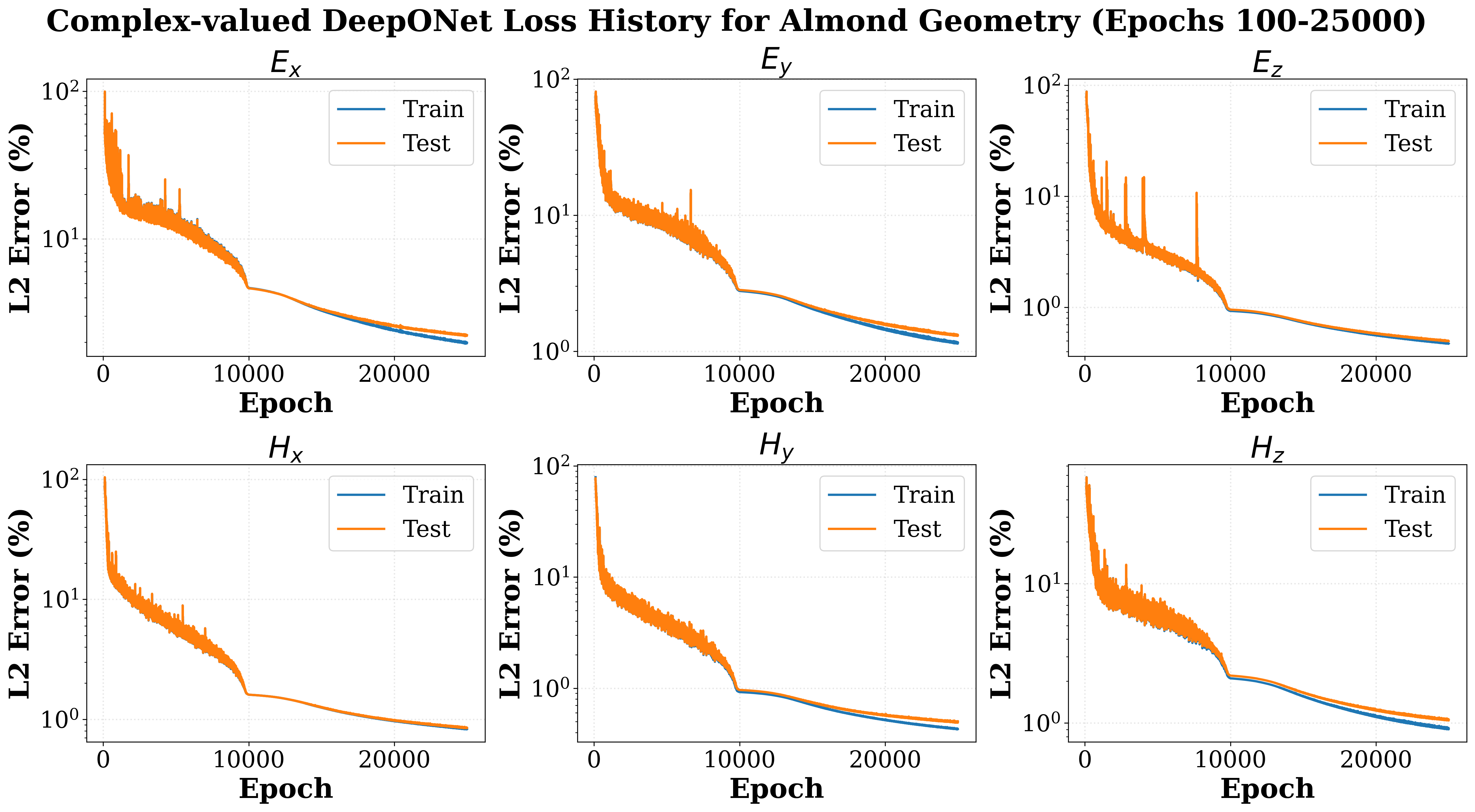}
    \caption{\textcolor{black}{Training and testing loss histories for the complex-valued DeepONet for the 3D Almond shape geometry shown in \autoref{fig:almond_geometry}. The loss shows show convergence for all six field components ($E_x$, $E_y$, $E_z$, $H_x$, $H_y$, $H_z$).The training strategy adopted the same early stopping criteria as in the previous experiment to avoid the overfitting.} }
    \label{fig:almond_loss}
\end{figure}

\textcolor{black}{The training loss histories for all six field components are shown in \autoref{fig:almond_loss}. The final relative $L_2$ errors are summarized in \autoref{almond_l2}. The training errors range from 0.15\% to 0.71\%, while the testing errors range from 0.22\% to 1.24\%. These low error values indicate that the complex-valued DeepONet generalizes well to unseen incident angles even on this geometrically complex target.}

\begin{table}[h!]
\centering
\caption{\textcolor{black}{Relative $L_2$ errors for training and testing of the complex-valued DeepONet on the Almond geometry at 1.19 GHz}} \label{almond_l2}
\vspace{1mm}
\begin{tabular}{|c|c|c|}
\hline
\textbf{Component} & \textbf{Training $L_2$ \%} & \textbf{Testing $L_2$ \% } \\
\hline
$E_x$& 0.71 & 1.24 \\
\hline
$E_y$& 0.41 & 0.68 \\
\hline
$E_z$& 0.16 & 0.23 \\
\hline
$H_x$& 0.31 & 0.37 \\
\hline
$H_y$& 0.15 & 0.27 \\
\hline
$H_z$& 0.31 & 0.59 \\
\hline
\end{tabular}
\end{table}

\begin{figure}[h!]
    \centering
    \includegraphics[width=0.95\textwidth]{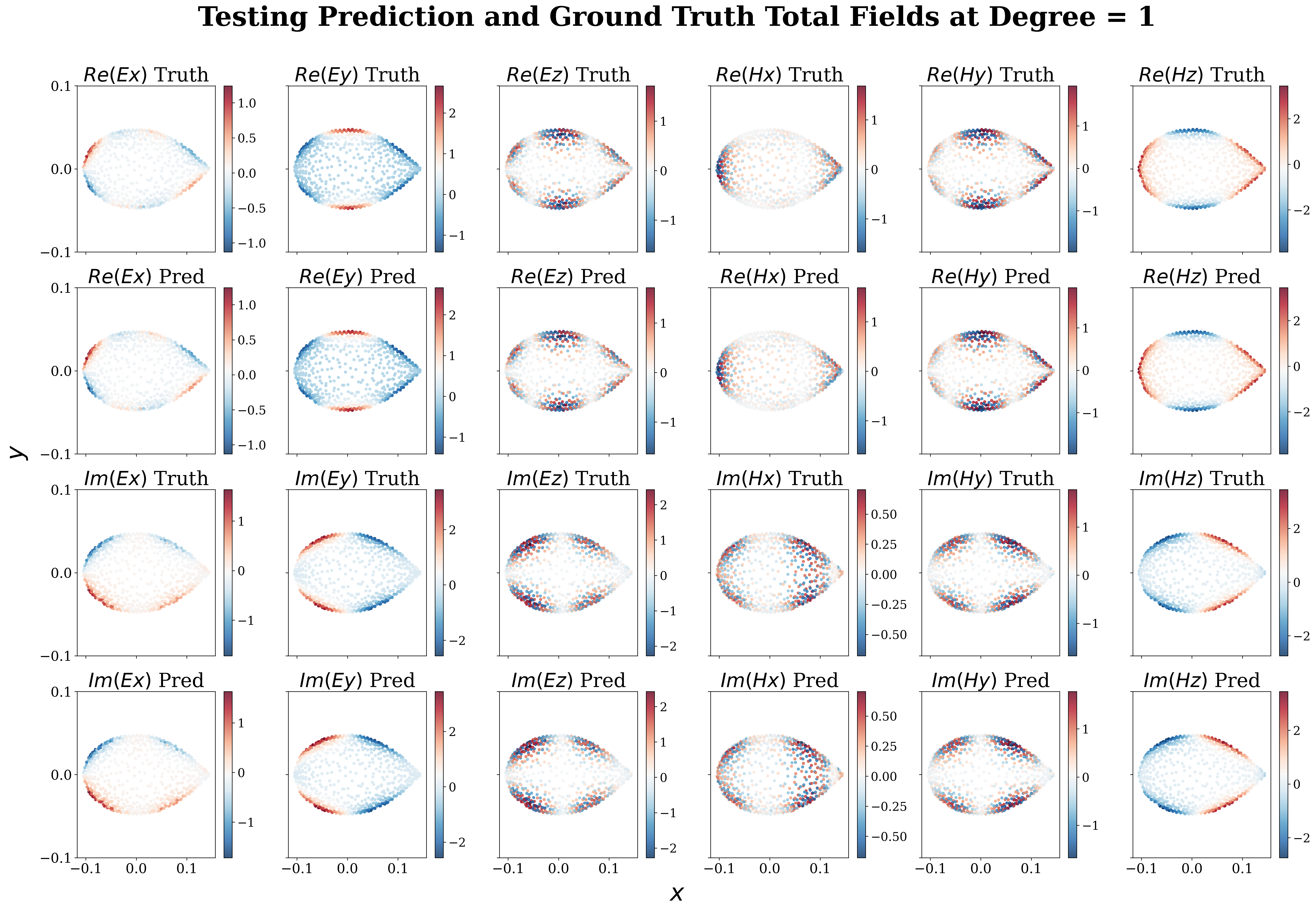}
    \caption{\textcolor{black}{Comparison between ground truth and predictions of the electromagnetic fields on the Almond geometry shown in \autoref{fig:almond_geometry} for a testing incident angle at $\phi = 1^{\circ}$. The top four rows show the real and imaginary parts of the electric field components (top two rows) and magnetic field components (bottom two rows).}}
    \label{fig:almond_predictions}
\end{figure}

\begin{figure}[h!]
    \centering
    \includegraphics[width=0.95\textwidth]{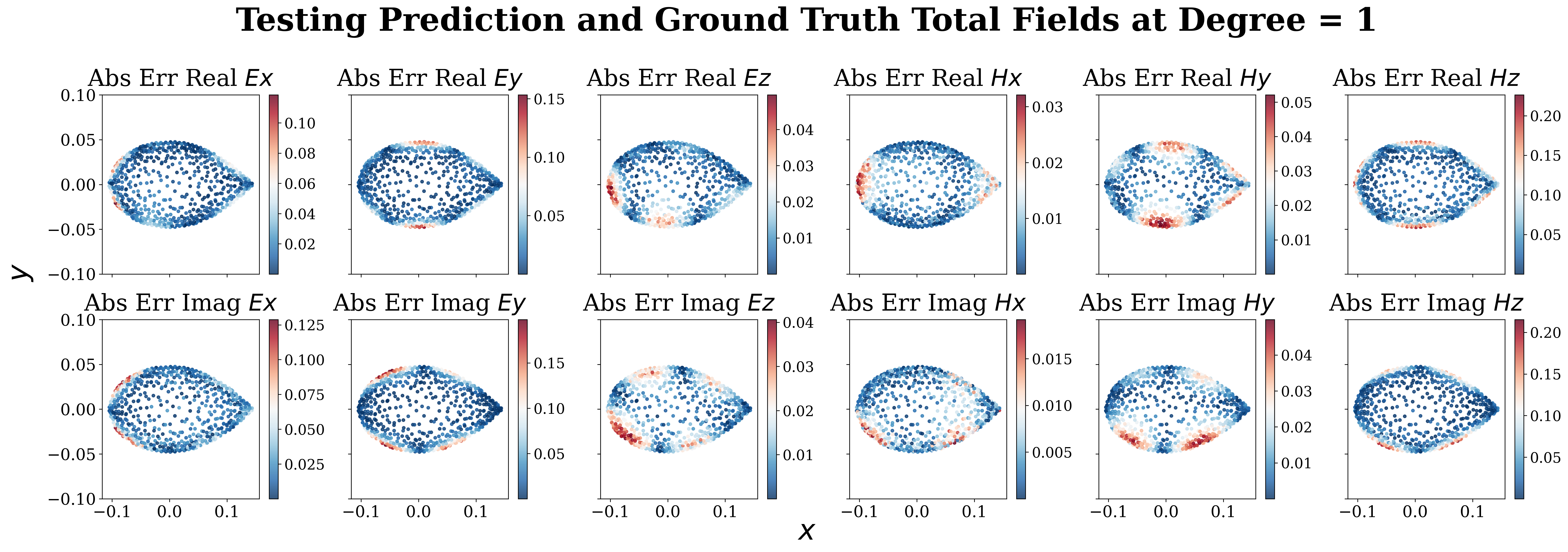}
    \caption{\textcolor{black}{Absolute pointwise errors for all six electromagnetic field components on the Almond geometry shown in \autoref{fig:almond_geometry} at a testing incident angle of $\phi = 1^{\circ}$. The errors are small throughout the domain, with slightly larger values concentrated near the sharp edges and corners where field singularities occur.}}
    \label{fig:almond_errors}
\end{figure}

\textcolor{black}{\autoref{fig:almond_predictions} presents a comparison between the ground truth and predicted fields for a testing incident angle at $\phi = 1^{\circ}$. The top two rows show the real and imaginary parts of the electric field components, while the bottom two rows show the magnetic field components. \autoref{fig:almond_errors} shows the absolute pointwise errors for all six components. The errors are generally very small across the computational domain, with larger errors concentrated primarily near the sharp edges and corners of the Almond surface.}

\textcolor{black}{These results demonstrate that the complex-valued DeepONet is capable of predicting electromagnetic fields on geometries with sharp edges and corners. In addition, the model can be scaled up to address more complicated problems where more expressivity is needed}

\subsection{Acquisition functions for sampling input frequencies}\label{sec:acq_fn_Results}

The acquisition function is used to efficiently select training frequencies for the complex-valued DeepONet. We consider the frequency range of $[0.05, 0.6] \, \text{GHz}$. An initial set of 5 uniformly distributed frequencies in this range are selected for training an ensemble of 3 DeepONets. Then, an additional 15 frequencies are acquired to maximize information gain as determined by the acquisition function. The trained model is subsequently tested on 80 unseen samples. 

\begin{figure}[h!]
    \centering
    \includegraphics[width=0.9\textwidth]{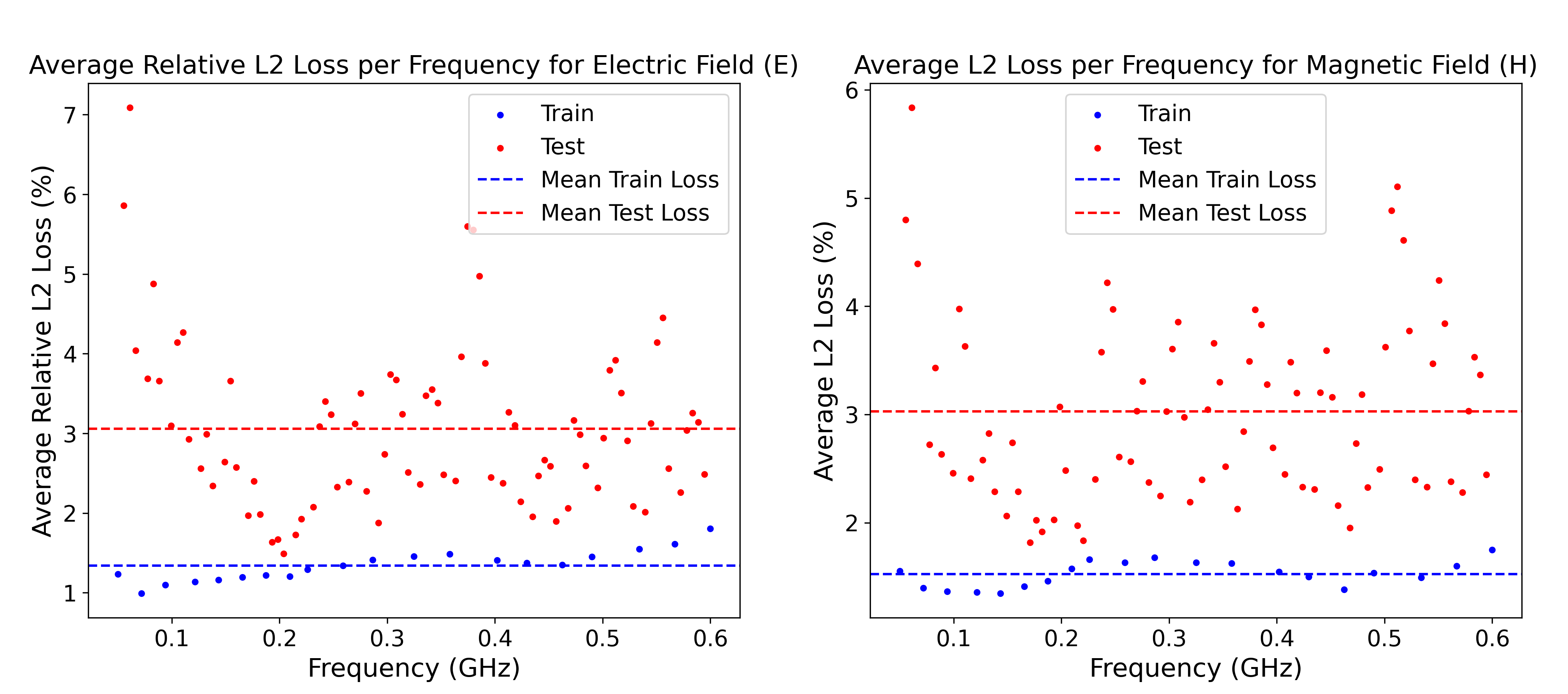}
    \caption{Scatter plots of relative $L^2$ losses for training (blue, 20 frequencies) and testing (red, 81 frequencies) selected using the acquisition function method from the frequency interval $[0.05, 0.6]$ GHz. The left plot shows average losses for electric field components ($E_x$, $E_y$, $E_z$), and the right for magnetic field components ($H_x$, $H_y$, $H_z$). Horizontal dotted lines mark the average losses for training and testing.}
    \label{fig:acq_loss_scatter}
\end{figure}

For a comparison, we also train the vanilla complex-valued DeepONet to randomly select 20 training samples, and both models are trained for 30000 epochs. The final testing losses in Figure \autoref{fig:acq_loss_scatter} indicate that the acquisition method significantly outperforms the original complex-valued DeepONet in predicting unseen testing data across all components of the electric and magnetic fields. \autoref{fig:acq_loss_scatter} illustrates the frequencies that are acquired for training and testing with their respective relative $L^2$ losses, which shows that with a very limited training dataset, our model can still generalize well to unseen testing data. 

\begin{figure}[h!]
    \centering
    \includegraphics[width=0.45\textwidth]{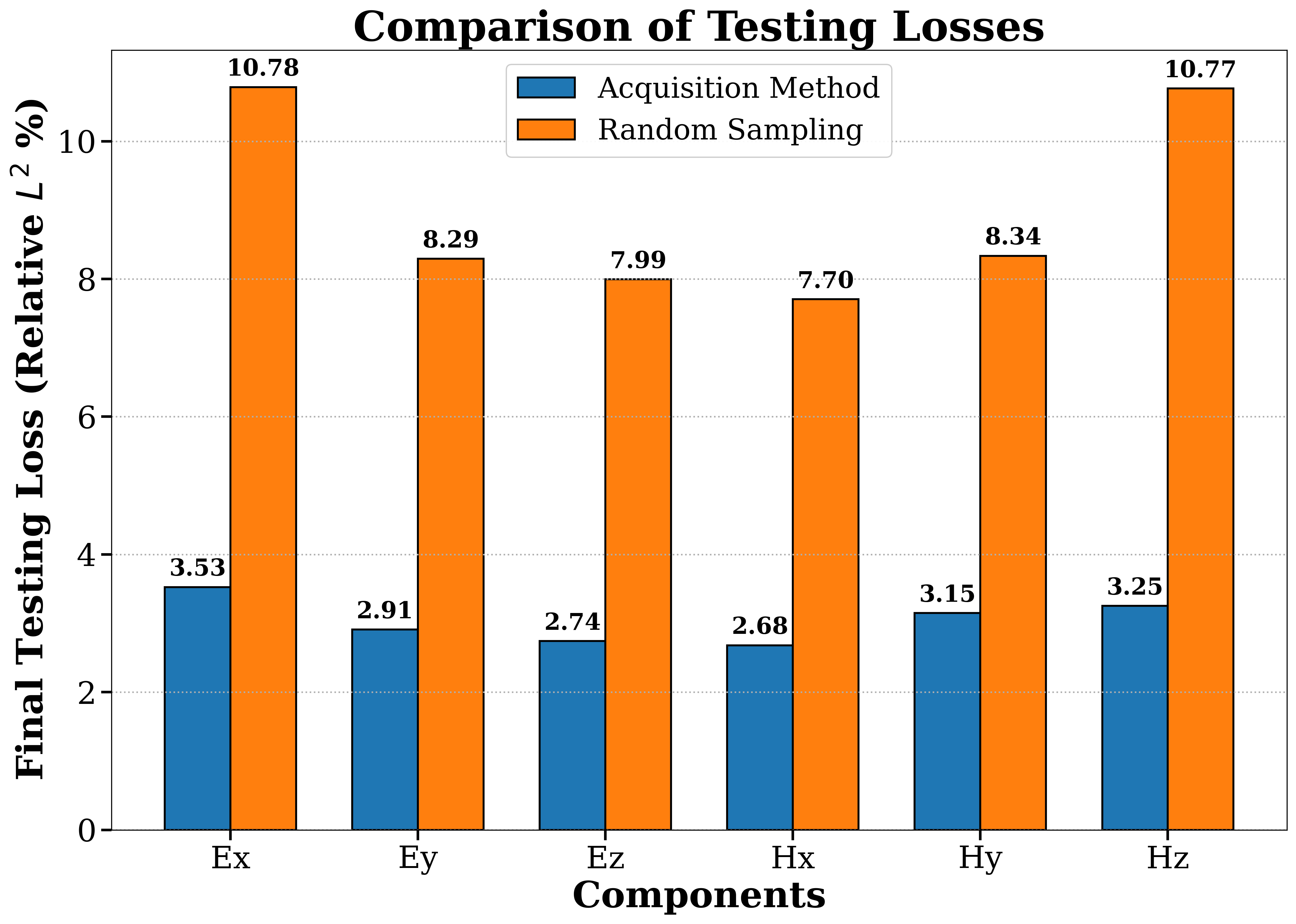}
    \caption{Comparison of relative $L^2$ errors in testing predictions for each component of the electromagnetic field using two sampling methods: the acquisition function sampling method and the random sampling method. Both methods are applied to the vanilla complex-valued DeepONet, trained with 20 data samples and tested with 81 samples.}
    \label{fig:compare_acq_vs_vanilla}
\end{figure}

\subsection{DeepONet as surrogate for computing radar cross section (RCS) results}\label{sec:RCS}
The overarching goal of the proposed DeepONet architecture is to efficiently compute integrated quantities, such as the radar cross section (RCS), which is critical for detecting and characterizing the signatures of different objects in real time. To demonstrate this capability, we compute the RCS using the electromagnetic fields inferred from the proposed DeepONet and compare the results against those obtained from a conventional numerical solver. This comparison allows us to assess both the accuracy and reliability of the DeepONet predictions in capturing essential physical quantities, while highlighting the potential of the approach for real-time applications where rapid evaluation of object signatures is required. Another important motivation for this exercise is to assess the sensitivity of integrated quantities to errors in the inferred fields. Even small inaccuracies in the predicted electromagnetic fields can propagate and result in significantly larger errors in computed integrated quantities such as the radar cross section (RCS). This sensitivity analysis is particularly important because many existing works proposing surrogate models often overlook this aspect, focusing primarily on pointwise field accuracy without evaluating its impact on derived quantities. 

The radar cross section (RCS) of a target is an equivalent area that reflects the transmitted power in a given direction. In many applications, the RCS derived from far-field patterns of the scattered wave is often the quantity of interest. The electric far-field pattern $\boldsymbol{E}_{\infty}$ is defined in spherical coordinates $(r, \theta, \phi)$ as 
\begin{align}
    \boldsymbol{E}(r, \theta, \phi) \approx \boldsymbol{E}_{\infty}(\theta, \phi) \frac{e^{i \omega \sqrt{\mu \epsilon} r}}{r},
\end{align}
when $r \to \infty$ \cite{Huttunen_2007}. 

\begin{figure}[h!]
    \centering
    \includegraphics[width=0.8\linewidth]{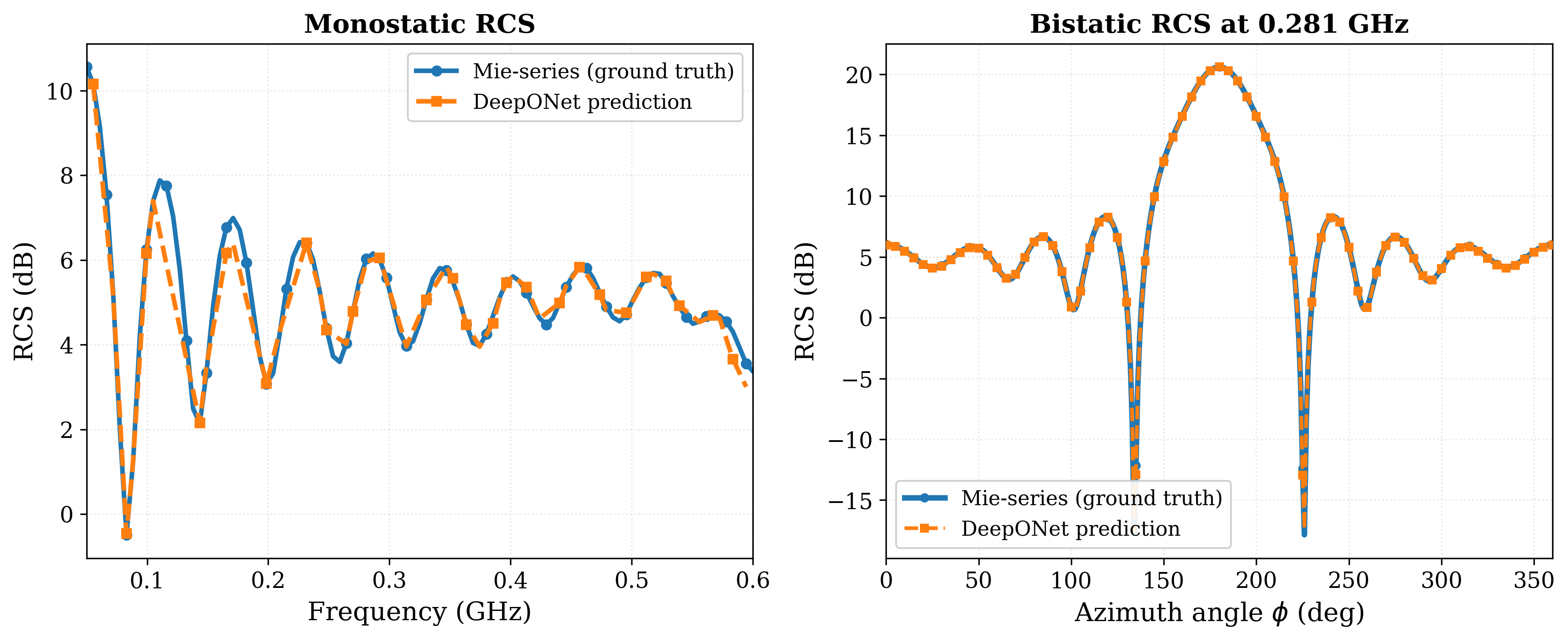}
    \caption{Monostatic and bistatic RCS results for the spherical scatterer, shown in \autoref{fig:geometry}, computed from DeepONet predictions (orange) and compared to ground-truth Mie-series solutions (blue). The monostatic RCS (left) is evaluated at 40 testing frequencies from the interval $[0.05, 0.6]$ GHz. The bistatic RCS (right) is calculated at a single frequency of $0.281$ GHz and plotted against the azimuth angle $\phi$. Predictions (dashed lines with square markers) match the ground truth (solid lines with circle markers) with high accuracy.}
    \label{fig:RCS_plot}
\end{figure}

The RCS is expressed as the ratio of the power density of the scattered field evaluated in the far-field to the incident power density times the area of a sphere of radius $r$, in the limit $r \to  \infty$. The monostatic and bistatic radar cross sections (RCS), computed using field predictions from the complex-valued DeepONet, are presented in \autoref{fig:RCS_plot}. For these results, the DeepONet was trained on 60 samples and tested on 40 samples. In \autoref{fig:RCS_plot}, the monostatic RCS is calculated at an azimuth angle of $\phi = 0$ for frequencies ranging from 0.05 to 0.6 GHz, while the bistatic RCS is shown for a frequency of $0.281$ GHz with $\phi$ varying from $0$ to $360$ degrees. For both monostatic and bistatic RCS calculations, the analytical Mie-series solutions are used as the ground truth for comparison. The results of RCS ensures the accuracy proposed DeepONet model for computing the integrated quantity, which is very sensitive to the accuracy of predict electromagnetic field from DeepONets.

{
The monostatic RCS for the Almond target is presented in \autoref{fig:RCS_almond}. The DeepONet was trained on 30 incident angles and tested on 31 unseen angles at a frequency of 1.19 GHz. \autoref{fig:RCS_almond} shows that the predicted RCS values (orange squares) agree well with the ground truth UWVF solutions (blue line). 

\begin{figure}[h!]
    \centering
    \includegraphics[width=0.6\linewidth]{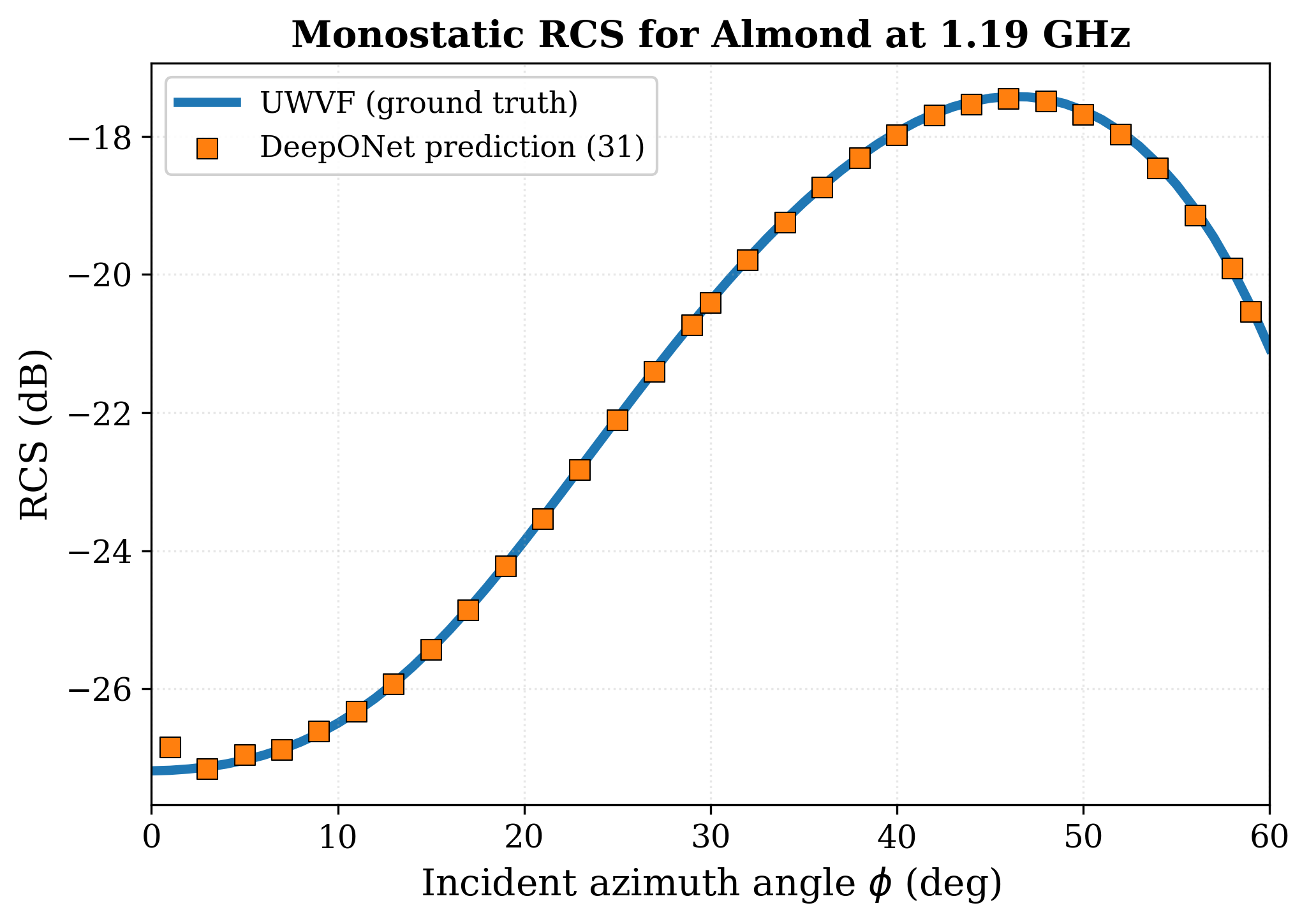}
    \caption{Monostatic RCS results for the Almond scatterer, shown in \autoref{fig:almond_geometry}, at 1.19 GHz as a function of incident azimuth angle $\phi$. 
    The ground truth UWVF solution (blue solid line) is compared with DeepONet predictions (orange squares, 31 testing angles). The predictions agree well with the ground truth across the angular range} 
    \label{fig:RCS_almond}
\end{figure}

}

\section{Summary}\label{sec:Summary}

In this study, we have developed a novel complex-valued DeepONet to address the challenge of operator learning for solutions expressed by complex  numbers. This approach preserves the phase information of complex numbers, which offers a significant advantage over real-valued models for problems governed by inherently complex-valued equations. This study leads to important contributions that enhance the efficiency and applicability of neural operators for solving direct scattering problems and computing the radar cross section governed by the time-harmonic Maxwell's equations. 

\textcolor{black}{ This work presents detailed numerical examples demonstrating the effectiveness of the complex-valued DeepONet in predicting the total electromagnetic fields from inputs consisting of the incident fields and the angular frequency. We consider two three-dimensional geometries: a metallic sphere with a smooth surface and a more complex almond-shaped domain featuring sharp corners. In both cases, the proposed complex-valued model outperforms the conventional real-valued DeepONet in terms of prediction accuracy and maintains robust performance when scaled to larger datasets. Furthermore, by incorporating an acquisition function to strategically sample the frequency input space, the model is able to achieve high accuracy even with a very limited number of training samples, highlighting its efficiency and potential for real-time applications.}

This study not only extends the DeepONet for complex-valued data but also introduces a technique to combine multiple DeepONet models to impose the boundary conditions for PEC scatterers. We have additionally adapted \textcolor{black}{the} reparametrization and two-step training method of real-valued DeepONet to the complex valued DeepONet to enhance the efficiency of model training and accuracy. 

This work sets the foundation for applying operator learning techniques to other complex-valued PDEs, e.g. Helmholtz equation, Schrodinger equations etc. A direction for further work would be to apply the method to model more complicated systems and reduce the computational cost of conventional numerical methods. 

\section*{Acknowledgments}\label{sec:Acknowledgements}
We gratefully acknowledge the financial support from Hypercomp Inc. This research was partially conducted using computational resources and services at the Center for Computation and Visualization (OSCAR) at Brown University. Additionally, we thank the AFOSR Grant FA9550-23-1-0671 for supporting the purchase of modern GPU hardware. Finally, we thank the valuable insights and support from Prof. George Em Karniadakis at Brown University.

\section*{Data and code availability}
The data and code related to the work will be made available at the GitHub repository \href{https://github.com/Qile-J/CV_DeepONet.git}{https://github.com/Qile-J/CV\_DeepONet.git}.


\appendix

\section{Comparison between Complex-Valued and Real-Valued Neural Networks}\label{appendix:cvnn_comparison}

\textcolor{black}{To illustrate why the complex-valued neural networks (CVNNs) are better than real-valued neural networks for learning inherently complex-valued functions, we present a simple one-layer network comparison. We consider the function $f(z) = |z| e^{i k \arg(z)}$ with $k=5$ and $z \in \mathbb{C}$.}

\subsection{Network architecture} \label{appendix:network_architecture}

\textcolor{black}{The complex-valued neural network used in this example is a standard one-hidden-layer network, where all operations exist in the complex domain $\mathbb{C}$. The NN-approximated function $f_{\text{complex}}(z)$ is given as:}
\begin{align}
\textcolor{black}{f_{\text{complex}}(z) = \mathbf{w}^{(2)} \cdot g_{\mathbb{C}}(\mathbf{W}^{(1)} z + \mathbf{b}^{(1)}) + b^{(2)}}
\end{align}
\textcolor{black}{where the input is $z \in \mathbb{C}$, the hidden layer has $\mathbf{W}^{(1)} \in \mathbb{C}^{H \times 1}$ (complex weights with $H$ hidden units), $\mathbf{b}^{(1)} \in \mathbb{C}^{H}$ (complex biases), and $g_{\mathbb{C}}: \mathbb{C} \to \mathbb{C}$ is a complex activation function as defined in \autoref{eq:activation_fn}. The output layer has $\mathbf{w}^{(2)} \in \mathbb{C}^{1 \times H}$ and $b^{(2)} \in \mathbb{C}$.}

\textcolor{black}{The real-valued approach uses two completely separate networks that do not share any weights or information. The final function $f_{\text{real}}(z)$ combines two independent outputs for the real part $u$ and imaginary part $v$:}
\begin{align}
\textcolor{black}{f_{\text{real}}(z) = u(x, y) + i \cdot v(x, y)}
\end{align}
\textcolor{black}{where the input is $\mathbf{x} = \begin{pmatrix} x \\ y \end{pmatrix} \in \mathbb{R}^2$ (the real and imaginary parts of $z$ treated as separate features). The real part network is:}
\begin{align}
\textcolor{black}{u(x, y) = \mathbf{w}_u^{(2)} \cdot g_{\mathbb{R}}(\mathbf{W}_u^{(1)} \mathbf{x} + \mathbf{b}_u^{(1)}) + b_u^{(2)}}
\end{align}
\textcolor{black}{and the imaginary part network is:}
\begin{align}
\textcolor{black}{v(x, y) = \mathbf{w}_v^{(2)} \cdot g_{\mathbb{R}}(\mathbf{W}_v^{(1)} \mathbf{x} + \mathbf{b}_v^{(1)}) + b_v^{(2)}}
\end{align}
\textcolor{black}{where all weights and biases are real-valued ($\in \mathbb{R}$), and $g_{\mathbb{R}}$ is ReLU.}

\subsection{Why the Complex Network Performs Better}

\textcolor{black}{The function $f(z) = |z| e^{i 5 \arg(z)}$ is inherently polar. It takes $z = r e^{i\theta}$ and maps it to $r e^{i(5\theta)}$, rotating the phase by 5x while keeping the magnitude. The core operation in a complex dense layer, $W \cdot z$, naturally performs rotation and scaling:}
\begin{align}
\textcolor{black}{W \cdot z = (|W| \cdot |z|) \cdot e^{i (\arg(W) + \arg(z))}}
\end{align}
\textcolor{black}{The network's basic operation adds phases (rotates) and multiplies magnitudes. This aligns well with the structure of the target function.}

\textcolor{black}{In contrast, the real-valued networks do not operate on angles or magnitudes directly. They process $x$ and $y$ coordinates separately. The real part network must learn:}
\begin{align}
\textcolor{black}{u(x, y) = \sqrt{x^2 + y^2} \cdot \cos(5 \cdot \text{atan2}(y, x))}
\end{align} 
\textcolor{black}{and the imaginary part network must learn:}
\begin{align}
\textcolor{black}{v(x, y) = \sqrt{x^2 + y^2} \cdot \sin(5 \cdot \text{atan2}(y, x))}
\end{align}
\textcolor{black}{These are complex, periodic, trigonometric functions. Approximating them with piecewise-linear ReLU activations is difficult. Furthermore, the two networks cannot share information. Each must independently learn $\sqrt{x^2 + y^2}$ and $\text{atan2}(y, x)$ without coordination.}

\subsection{Experimental Results}

\begin{figure}[h!]
    \centering
    \includegraphics[width=0.9\textwidth]{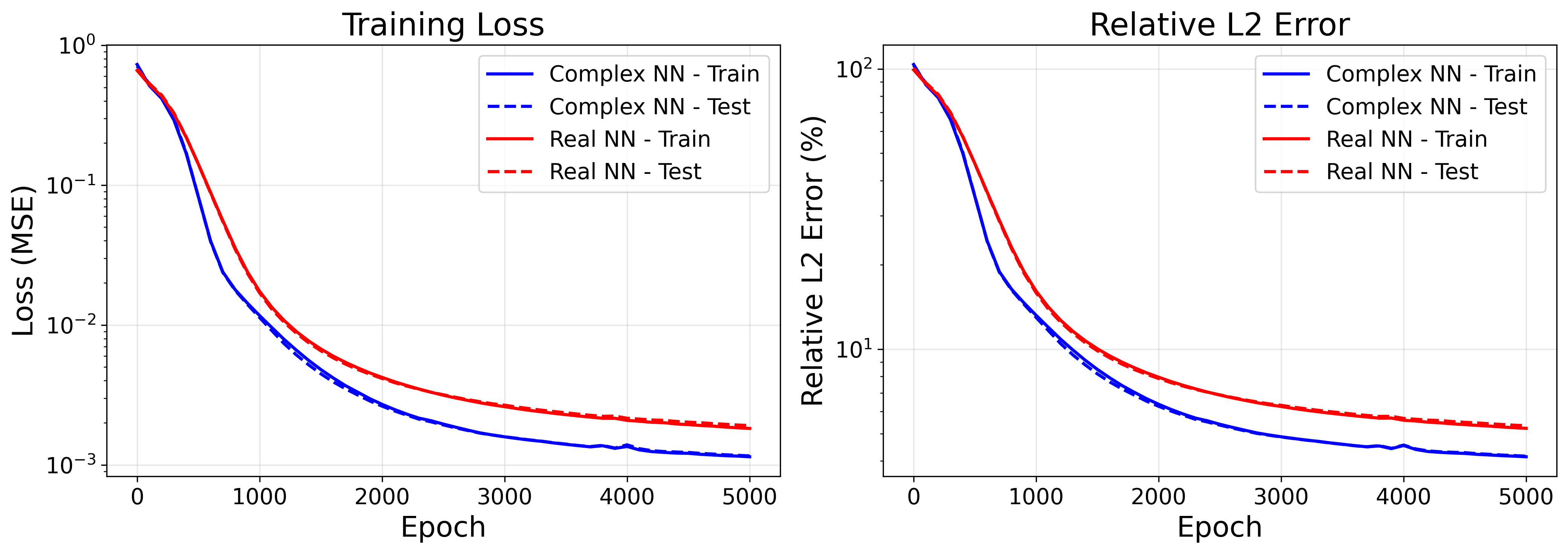}
    \caption{\textcolor{black}{Training and testing loss histories for the complex-valued neural network (blue) and real-valued neural networks (red). The left plot shows the mean squared error loss, while the right plot shows the relative L2 error. The complex-valued network converges to a lower error.}}
    \label{fig:cvnn_loss_comparison}
\end{figure}

\textcolor{black}{We generate 10,000 training samples and 2,000 test samples by randomly sampling complex numbers from $[-1,1] + i[-1,1]$. Both models use only one hidden layer with 64 neurons and are trained for 5,000 epochs with the Adam optimizer at learning rate $10^{-3}$. The training and testing loss histories are shown in \autoref{fig:cvnn_loss_comparison}. The complex-valued network achieves a final test error of 4.16\%, while the real-valued approach achieves 5.34\%.}

\begin{figure}[h!]
    \centering
    \includegraphics[width=0.95\textwidth]{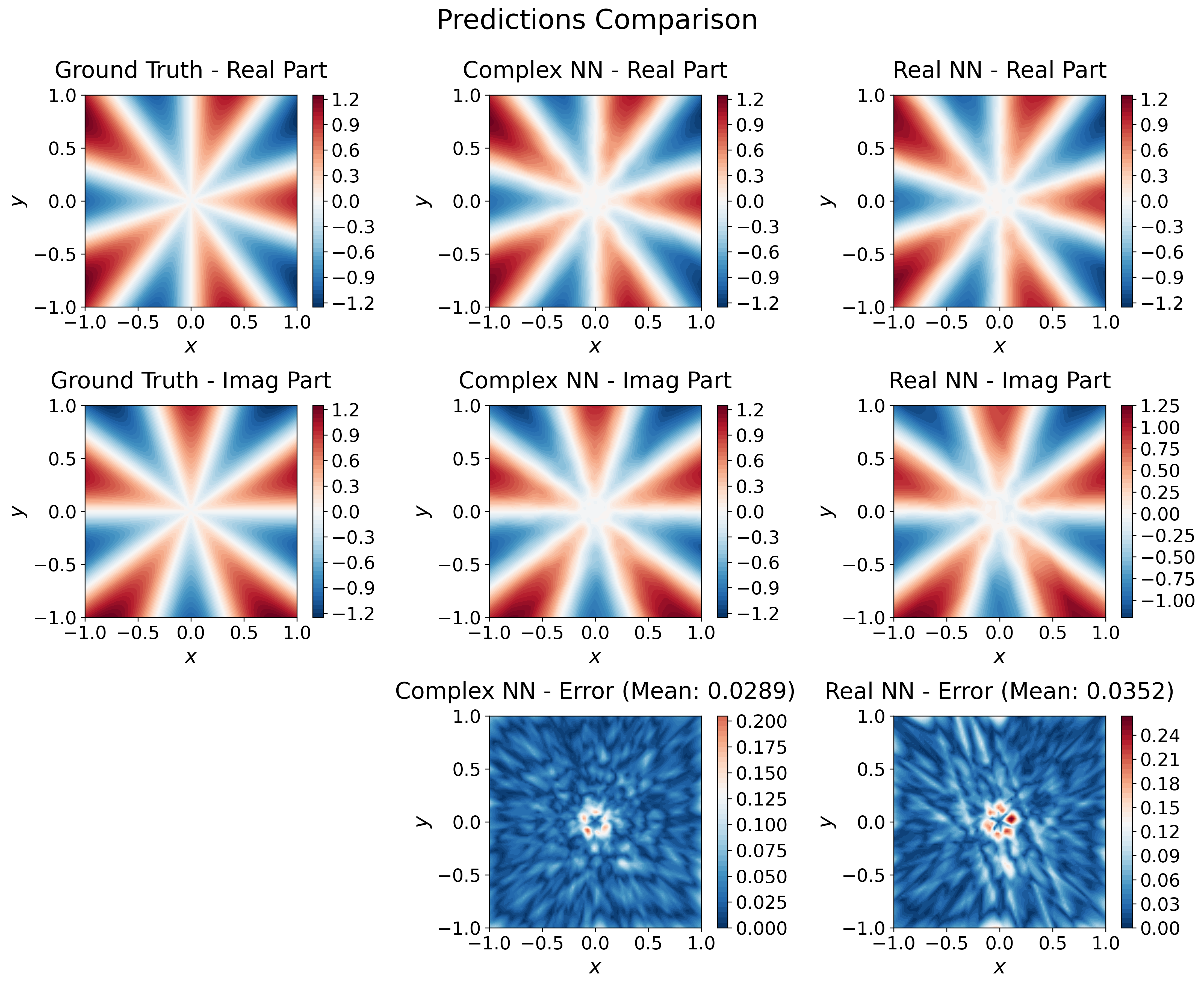}
    \caption{\textcolor{black}{Comparison of predictions for $f(z) = |z| e^{i 5 \arg(z)}$. The complex-valued network captures the intricate patterns around the origin more accurately than the real-valued approach.}}
    \label{fig:cvnn_predictions_comparison}
\end{figure}

\textcolor{black}{\autoref{fig:cvnn_predictions_comparison} presents predictions on a uniform grid. The top row shows real parts, the middle row shows imaginary parts, and the bottom row shows absolute pointwise errors. The complex-valued network (mean error: 0.0289) achieves lower errors than the real-valued approach (mean error: 0.0352).}

\textcolor{black}{This example suggests that for problems with inherent phase and magnitude structure, complex-valued neural networks may offer advantages. This motivates the design of complex-valued DeepONets to solve PDEs with complex-valued solutions.}


\bibliographystyle{unsrt}
\bibliography{references}  

\end{document}